\documentclass[12pt]{iopart}
\pdfoutput=1

\usepackage{citesort}

\usepackage[square,sort&compress,comma,numbers]{natbib}

\usepackage[breaklinks,colorlinks=true]{hyperref}
\usepackage{iopams}
\usepackage{graphicx}
\usepackage{appendix}

\begin{document}


\title[nesmcq18-TASEP+defect]{Passive Tracer Dynamics in Slow-Bond Problem}

\author{Hyungjoon Soh\footnote{Present address: Kakao Brain, Pangyoyeok-ro 241, Seongnam, 13494, Korea}}
\address{Department of Physics, Korea Advanced Institute of Science and Technology, Daejeon
34141, Korea}
\author{Meesoon Ha\footnote{Author to whom any correspondence should be addressed.}}
\address{Department of Physics Education, Chosun University, Gwangju 61452, Korea}
\ead{msha@chosun.ac.kr}
\date{\today}

\begin{abstract}
\noindent
Asymptotic Kardar-Parisi-Zhang (KPZ) properties are investigated in the totally asymmetric simple exclusion process (TASEP) with a localized geometric defect. In particular, we focus on the universal nature of nonequilibrium steady states of the modified TASEP. Since the original TASEP belongs to the KPZ universality class, it is mathematically and physically a quite interesting question whether the localized columnar defect, the slow bond (SB), is really always relevant to the KPZ universality or not. However, it is numerically controversial to address the possibility of the non-queued SB phase in the weak-strength SB limit. Based on the detailed statistical analysis of KPZ-type growing interfaces, we present a comprehensive view of the non-queue SB phase, compared to finite-size crossover effects that reported in our earlier work [Soh {\it et al.}, Phys. Rev. E {\bf 95}, 042123 (2017)]. Moreover, we employ two types of passive tracer dynamics as the probe of the SB dynamics. Finally, we provide intuitive arguments for additional clues to resolve the controversy of the SB problem.  
\end{abstract}

\\

\noindent\textbf{ArXiv}: 1905.00083



\maketitle

\tableofcontents 

\section{Introduction}
\label{sec:intro}

In nature, most of dynamic phenomena are far from equilibrium. While the theory of equilibrium systems has been well built, nonequilibrium systems do not have theoretically successful descriptions. One of the breakthroughs came out from the nonequilibrium interface growth of the Kardar-Parisi-Zhang (KPZ) equation\cite{Kardar1986}, described by a nonlinear stochastic differential equation. The KPZ equation represents, not only limited to interface growth such as body-centered solid-on-solid (BCSOS) growth~\cite{Meakin1986,Krug2000}, but also spans models described by the stochastic heat equation with multiplicative noise, such as directed polymers in random media (DPRM)~\cite{Kardar1987}, the directed last passage percolation (DLPP)~\cite{Johansson2000,Bodineau2005}, and so on.

Based on the lowest relevant orders and symmetries, it is known that the KPZ equation describes the wide range of nonequilibrium models, where the model details are different, but the macroscopic property among them coincides with one another. Although some subtle issues still left in questions, most of computational model tests and experimental studies have successfully confirmed the KPZ equation that describes in such models~\cite{Takeuchi2010,Takeuchi2011}. 

In a $(1+1)$-dimensional (1D) KPZ system, it is the simplest implementation of the broken translational invariance to add a single-site defect on the 1D space. The KPZ universality systems can be also considered with global defects under general spatial dimensions, such as in $(d+1)$-dimensional DPRM with $d$-dimensional planar attractive potentials, interface growth with random defects, stochastic transport with blockages, and the pivot enhancement in the DLPP. 

In general, the effect of the defect is irrelevant in the $(d+1)$-dimensional system as far as the defect dimension $D_<$ is lower than $d$, which means that such a defect does not affect the scaling property of the system~\cite{Tang1993}. Hence, for the defect dimension $D_>$ that is larger than $d$, one expects that the defect effect becomes relevant to the system. In the same conmbox, we pose the question: {\it What happens if $D=d$?}. For the marginal case, it is marginally either relevant or irrelevant. 

If the line defect ($D=1$) is marginally relevant to 1D KPZ interface growth, it can significantly alter the KPZ universality property in Tracy-Widom scaling limits. Accordingly, for past decades, the question whether any arbitrarily small value of the defect can always destroy the KPZ universality has been in discussion. The same question has been asked and answered with controversial issues~\cite{Janowsky1994,MHa2003,Costin2012,Basu2014,Schmidt2015,Soh2017} in the totally asymmetric simple exclusion process (TASEP), which is the exact mapping of the BCSOS growth and another interpretation of the KPZ equation: At any arbitrarily small defect strength, is it possible that the microscopically localized slow-bond (SB) defect cannot affect the macroscopic behavior of the system, termed as the longstanding ``SB problem''~\cite{Janowsky1994,MHa2003,Costin2012,Basu2014,Schmidt2015,Soh2017}.
On the other hand, in real-world studies the geometrical modifications of the original TASEP have were widely studied from biological transport to traffic. Some examples may also include forked paths, random directed networks, unlimited capacitances, and Langmuir adsorption/desorption process\cite{Parmeggiani2004,Embley2009,Brankov2004,Neri2011,Basu2010}. Such spatial and dynamic deformation plays equivalent role of SB in the TASEP. Thus, the SB problem is a important question in both physical and practical aspects.

Regarding the localization of the defect effect on the TASEP, most studies~\cite{Tang1993, Janowsky1994, Balents1994, Foulaadvand2008, Kinzelbach1995, Hwa1995, Lassig1998, MHa2003, JHLee2009, Costin2012, Seppalainen2001, Kandel1992, Basu2014, Schmidt2015, Soh2017} employed both analytic and numeric methods. Analytic studies~\cite{Janowsky1994} involving the mean-field (MF) analysis always predict that the SB affects globally, irrespective of its strength. However, the analytic approach is useful in the SB problem due to the difficulty of calculating the average of localized quenched randomness, which is quite nontrivial and precarious. On the other side, earlier numerical studies~\cite{MHa2003, Kandel1992,JHLee2009} predict a phase transition at the finite SB strength. However, most recent work~\cite{Soh2017} has suggested that the non-queued SB phase is a crossover phenomenon in finite systems, not the thermodynamically stable phase.

In this paper, we revisit the controversy of the critical SB strength, below which the SB effect is confined locally, in the conmbox of the KPZ universality statistics and two types of passive tracer dynamics, respectively. Due to the difficulty of numerical studies caused by finite-size effects and boundary conditions as well as initial-setup issues, we present a comprehensive view of the non-queued SB phase in the conmbox of the systematic data analysis and provide intuitive arguments that support the existence of non-queued SB phase in the weak SB limit. 

More specifically, we investigate the TASEP with a SB, in terms of 1D BCSOS growth, where we systematically quantify the SB effect as performing extensive Monte-Carlo (MC) simulations. In the presence of the SB, most of analytic solutions become unstable due to the broken spatial symmetry. Addressing the existence of the non-queued SB phase, we define the following four observables: surface fluctuations in transient 1D BCSOS growth, and the distribution of passive tracer location in the steady-state limit. 

The rest of the paper is organized as follows. In Sec.~\ref{sec:model}, we show that the TASEP with a SB is exactly mapped onto the 1D BCSOS growth with a columnar defect and passive tracer dynamics. Based on physical quantities of interest, we state key questions and clarify possible implications. In Sec.~\ref{sec:numerics}, we present extensive numerical simulation results and discuss two different time regimes: In the transient regime, we discuss the height statistics of growing surfaces, while in the stationary regime we discuss how the TASEP measure is changed due to the SB. Finally, we summarize the controversial results with some remarks in Sec.~\ref{sec:conclusion}. 

\section{Model} 
\label{sec:model}

\subsection{TASEP with a SB}

Consider a one-dimensional (1D) lattice with $L$ sites and the total number of $N(\le  L)$ particles. Each site $x$ can be occupied by at most a particle, so the occupancy number at site $x$ is $n(x) \in \left\{0, 1\right\} $. At each time step, a site is selected at random. If the chosen site is occupied and its right nearest neighboring (NN) site is empty, then the particle at the chosen site, say $x$, jumps to the right NN site, $x+1$, with unit probability. This is just as the ordinary TASEP. We here add a localized defect on the TASEP, which is implemented by assigning a special bond in the middle of the system with a reduced hopping probability, $r(\le 1)$, namely the SB. It is noted that the Monte-Carlo (MC) simulation time is always updated on the random site selection by the increment of time, $1/L$, regardless of the success of the particle jump. 

Two ends of the 1D lattice can be connected to either each other as periodic boundary conditions (PBC), or particle reservoir as open boundary conditions (OBC). For the PBC, the particle density $\rho_0=N/L$ is fixed as the initial value, which becomes a control parameter of the system, while for the OBC, the particle entry (from the left of the leftmost site) and the particle exit (to the right of the rightmost site) are controlled with respect to each particle reservoir density:  
$\alpha\equiv\rho_{\rm left}$ and $\beta\equiv 1-\rho_{\rm right}$, where $\alpha$ is the probability that a particle from the left particle reservoir tries to enter to the leftmost site of the system and $\beta$ is the probability that a particle at the rightmost site of the system exits from the system.  


\subsection{Physical Quantities of Interest}

\subsubsection{Transient Regime}

The main purpose of this paper is to observe the true impact of the SB in the thermodynamic limit through various observables, in particular to the conmbox of the KPZ universality class. Since it is well known that the KPZ-type models exhibit distinctive features in both time and space, one can classify physical observables of interest for such models, based on the observation time and length scales. Our starting point is to focus on the transient regime before the system reaches the steady state.  Under the circumstances, KPZ fluctuations are far from equilibrium. In recent mathematical studies for the 1D KPZ universality class (see \cite{Corwin2011} and references therein), the rescaled height distribution as well as its fluctuations and correlation functions between heights, are exactly derived for six fundamental initial conditions.

We here limit the TASEP with the SB to start at the flat initial condition: 
$[h_0]_x = 0$ for 1D BCSOS growth and $\rho_0=[n_0]_x=1/2$ for the TASEP, where $[\cdot]_x$ is the spatial average. In the case of 1D BCSOS growing interface without the SB, the height function $h(x,t)$ at asymptotically large time is written rigorously as follows (see \cite{Ferrari2005, Sasamoto2005, Calabrese2011}):
\begin{equation}
\lim_{t\rightarrow\infty}\frac{t/2-h(x,t)}{t^{1/3}}(\equiv\tau)\longrightarrow^{d~} F_1(2\tau).
\end{equation}
Here the rescaled height distribution of ordinary KPZ fluctuations, $F_1(\tau)$, is independent of $x$. However, when the SB is inserted to the system, $h(x,t)$ is not symmetric any more in space, {\it i.e.}, spatially inhomogeneous. To investigate such a inhomogeneity caused by the SB strength, we measure height fluctuations with respect to the average with stochastic realizations for both the conventional definition $W^2$ and the site-dependent one $\sigma_w^2$. 

In presence of the SB, there is no well-defined form for the deterministic part of $h(x,t)$ to define $\tau$ properly. This is why we directly investigate the cumulant ratios for such unknown distributions. More specifically, we focus on the standard deviation $\sigma_w$, the skewness $S$, and the kurtosis $K$ for height profiles as a function of space and time, which are defined as follows:
\begin{eqnarray}
\sigma_w^2(x,t) & = & \langle \left(\delta h(x,t)\right)^2\rangle,
\label{eq:sigma}\\
S(x, t) & = & \langle \left(\delta h(x,t)\right)^3\rangle/\sigma_w^{3}(x,t),
\label{eq:S}\\
K(x, t) & = & \langle \left(\delta h(x,t)\right)^4\rangle/\sigma_w^{4}(x,t)-3,
\label{eq:K}
\end{eqnarray}
where we denote $\delta h(x,t)\equiv h(x,t)-\langle h(x,t)\rangle$ and 
$\langle\cdot\rangle$ for the ensemble average. Note that the standard deviation $\sigma_w$ is distinguished from the conventional surface width $W$ for homogeneous growth models,
\begin{eqnarray}
W^2(L, t) = \left[\langle \left(h(x,t)-[h(x,t)]_x\right)^2\rangle\right]_x.
\end{eqnarray}
Here $[f(x)]_x\equiv\sum_{x=1}^{L} f(x)/L$ is independent of $x$.

Starting with 1D BCSOS growth with the flat initial condition, the alternatively ordered configuration of the TASEP is broken at the early stage of MC simulations by the random deposition of particles. In this regime, the exclusion does not occur dominant, so that the surface fluctuates as Gaussian. As sufficiently long time elapsed, the slowest order of fluctuations governs global fluctuations, namely KPZ fluctuations. Simultaneously, the SB builds up to the global facet and spreads to the whole system, which eventually distorts KPZ fluctuations. The detailed discussion will be provided in Sec.~\ref{sec:numerics}.

\subsubsection{Steady-State Regime}
 
In the steady-state limit when both spatio-temporal correlations develop as time goes by and eventually cover the whole system, we measure time-independent properties of physical quantities. Particle configurations in the TASEP keep fluctuating around the average density profile $\rho(x)=\langle n(x)\rangle$, while 1D BCSOS growing surface configurations fluctuate around the average height profile $\langle h(x)\rangle$. Since the stationary current $J$ in the TASEP is a well-defined quantity, which corresponds to the growing velocity $v$ in BCSOS surface growth, it is useful to check the fundamental relation between $J$ and the bulk density $\rho_b$ as the SB strength varies. 

For the TASEP in the absence of the SB, 
$$J_{_{\rm OBC}} = \min\left[\alpha(1-\alpha), \beta(1-\beta), 1/4\right],$$
$\rho_b=\min[\alpha,1/2]$ (low-density);  
$\max[1-\beta, 1/2]$ (high-density);  1/2 (maximal-current) for the open case, and 
$$J_{_{\rm PBC}} = \rho_b (1-\rho_b),$$ 
$\rho_b=\rho_0$ for the closed case.

The difficulty of quantifying the SB effect on the TASEP through numerical simulations arises from the competition between boundaries and the local defect in similar orders of magnitude. Moreover, in the aspect of the global $J$ calculation, finite-size corrections should be considered, which depend on the boundary conditions and obey the following analytic form~\cite{Derrida1993,Blythe2007}:
\begin{eqnarray*}
J_{_{\rm OBC}}\left(L, \alpha+\beta =1 \right) &=& J_{_{\rm OBC}}\left(L = \infty\right), \\
J_{_{\rm OBC}}\left(L, \alpha=\beta =1 \right) &= &J_{_{\rm OBC}}\left(L = \infty\right) + \frac{3}{8L} + O\left(L^{-2} \right), \\
J_{_{\rm PBC}}\left(L, \rho \right) &=& J_{_{\rm PBC}}\left(L = \infty\right) + \rho\left(1-\rho\right)\frac{1}{L-1}.
\end{eqnarray*} 
As a result, we find that the open TASEP without the SB has the special line ($\alpha+\beta=1$), where no finite-size corrections exist in the global current. This is why we choose $\alpha=\beta=1/2$ for the OBC.
Moreover, the SB problem and its variant were also studied with a special site of the unlimited capacity in the TASEP with the PBC, namely the parking garage model~\cite{MHa2002}. Since the SB itself behaves as another boundary, which controls $J$ but preserves the fluctuations of $\rho$. In contrast, the OBC do not conserves $\rho$ directly. According to the capacity of the parking garage, one can easily consider both the open and the closed case. 

Discussing the ensemble equivalence with the direct comparison of the OBC with the PBC, physically interesting quantities are expressed in terms of the height function $h(x,t)$ that suffices the KPZ equation as follows: 
\begin{eqnarray} 
\label{eq:KPZ+J+rho}
\displaystyle \frac{\partial h(x,t)}{\partial t} &=& \nu_h\nabla^2 h(x,t) + \frac{\lambda}{2}\left( \nabla h(x,t) \right)^2 + \eta(x,t),\\
J& =&  \left \langle \lim_{t\to\infty} \left [ \frac{\partial h(x,t)}{ \partial t} \right]_x \right \rangle, \\
\rho(x)& = &\left \langle \lim_{t\to\infty} \left( \frac{\partial h(x,t)}{\partial x}\right) \right \rangle, 
\end{eqnarray}
where $\nu_h$ and $\lambda$ are some constants, and $\eta(x,t)$ is a Gaussian white noise function with 
$\langle\eta(x,t)\eta(x',t')\rangle\propto\delta(x-x')\delta(t-t').$

Finally, we clarify the SB problem in the conmbox of the KPZ universality class by a tagged passive walker in the steady-state limit, which traces height configurations without interrupting the KPZ dynamics, namely a passive tracer. The feedback of steady-state fluctuations to this passive tracer, can be either positive or negative. The positive/advection (negative/anti-advection) tracer jumps right if the next site is vacant (occupied), and left if the site is occupied (vacant). The passive tracer is kind of the second-class particle introduced to track shock fronts in the TASEP~\cite{Derrida1993a}, and its scaling properties without the SB were discussed in several studies~\cite{Drossel2002,
CSChin2002,HKim2011,Ueda2015}. 

Consider a passive tracer with positive/negative feedback at $x_{\pm}(t)$ at time $t$ driven by the SB biased particle field and its dynamics can be written as follows: 
\begin{equation}
\frac{dx_{\pm}(t)}{dt} = s_{\pm}\frac{\partial h(y,t)}{\partial y}|_{y=x_{\pm}(t)} + \zeta(t), 
\label{eq:tracer}
\end{equation}
where $s_{\pm}=\pm1$ (positive/negative feedback) and $\zeta(t)$ represents a Gaussian white noise. 

\section{Numerical results} 
\label{sec:numerics}

In this section, we present numerical results with the comparison of conjectured analytical results and physical arguments and provide three different points of view in discussing the SB problem. 


Preforming MC simulations of the TASEP with the localized SB, we employ both rejection-free continuous-time algorithms and standard ones. Initially, particles are all set in alternating order, $\rho_{0}=1/2$. First, we make the list of active sites at which particles can always jump, with respect to particle configuration for the rejection-free selection. 
At each step, a site is randomly selected and removed from the list of active sites. The particle at the selected site hops to the right NN site, and the active site list is updated by adjacent site configurations. If the selected active site tries to jump the SB, then a random number is generated~\footnote{We used the Mersenne Twister as a pseudo-random number generator (PRNG), which is by far the most widely used general-purpose PRNG and of which name derives from the fact that its period length is chosen to be a Mersenne prime (see http://www.math.sci.hiroshima-u.ac.jp/\~~m-mat/MT/ARTICLES/earticles.html for the details).}. 

For the OBC, the enter (exit) event is also included in the active site list, only if the leftmost (rightmost) site of the system becomes vacant (occupied). If the enter (exit) event is picked, a particle enters (exits) with probability $\alpha$($\beta$). At every $t = 2^n (n \in \{0, 1, ...\})$ MCS, $h(x,t)$ and its statistics are measured until $t > 100L^{3/2}$ to ensure the system can reach the steady state. The stationary-state cutoff is determined by temporal correlations, which decay as $Ct^{-2/3}$. The coefficient is numerically checked using surface width that has a stable value at $r = 1~(\epsilon=0)$, which belongs to the KPZ universality class. To measure stationary quantities of interest with the better quality, we save configurations at every inter-sampling time $L^{3/2}$ to avoid temporal correlations in samples. 

Moreover, we limit our study to focus on the impact of the SB at $\alpha=\beta=1/2$, which only allows the flatness of the density profile in the maximal-current phase ($\alpha,\beta \ge 1$) with the bulk density $\rho_{\rm mc}=1/2$ and the maximal current $J_{\rm mc}=1/4$. 
Since the characteristics of the TASEP depends on boundary conditions, the considerate choice of boundaries is essential to separate the SB effect from that of BCs. As discussed in a lot of earlier studies for the OBC without a SB,  the physical quantities of interest in the steady-state limit, such as the bulk density $\rho$ and the stationary current $J$, have the analytic forms under the condition of $\alpha + \beta = 1$, where spatial correlations vanish, so that no finite-size effects are left~\cite{Derrida1993}. As a result, one can easily find that the only BCs that suffice both the product state, and the maximal-current phase is the setting of $\alpha = \beta = 1/2$, unlike the case of the most recent numerical study~\cite{Schmidt2015}, 
where they set $\alpha=\beta=1$. Here we like to note that for the regime of $\alpha,\beta > 1/2$, there are always power-law decaying parts in both ends of the density profile, so that the bulk is never flat.  For the PBC, the choice of $\rho_0 = 1/2$ satisfies both the maximal-current phase and the flatness of the density profile in average. 

\subsection{TASEP with the SB}
We start with the steady measure for the TASEP with the SB. As the system reaches its steady state, every site has the equal rate of flow (the stationary current) $J$. Even in presence of the SB, its impact in a system directly affects the density difference across the SB, but the stronger correlations between adjacent jumps maintain $J$ per every bond. As a result, the stationary density profile $\rho(x)$ is determined by $J$. 


The density profile can be directly analyzed for various SB strength as discussed in the previous works\cite{MHa2003,MHa2011,Schmidt2015,Soh2017}, where it is found that in the sufficiently large size,  the partial evidence of ``$\epsilon_c\ne0$" exists based on the following physical ansatz for $\rho(x)$ in the presence of the SB: The SB effect induces the special structure to the density profile in the steady-state limit, where the power-law correction term is conjectured for the jamming tail 
$\Delta(x)=2\rho(x)-1$.
\begin{equation}
\Delta(x) = \Delta_b + Ax^{-\nu},
\label{eq:DP}
\end{equation} 
where $\Delta_b=2\rho_b-1$ denotes the offset of $\rho_b$ from the value in the absence of the SB, the group velocity $v_g=\frac{\partial J}{\partial \rho_b}$, and $x$ is the distance from the SB. Note that there is a conjecture\cite{MHa2003} for the decay exponent $\nu$ converges to $1/3$ as $\epsilon_c>\epsilon\rightarrow 0$ as $\Delta_b\to 0$ from $\epsilon=\epsilon_c$, 
where the SB asymptotically plays a role as the OBC with effective enter/exit rates. However, finite-size effects become dominant, which induces the pseudo-critical strength $\epsilon_c(L)$ in both the PBC and the OBC. So we pose the main question whether the non-queued SB phase can be stable below $\epsilon\approx 0.2$ in the thermodynamic limit.

Recently, a numerical study~\cite{Schmidt2015} has claimed that the stable non-queued SB phase at non-zero SB strength ($\epsilon<\epsilon_c\ne 0$) is attributed to finite-size effects from improper setups for the OBC. Moreover, it was argued that numerical tests eventually confirm the analytic form of $J_s(\epsilon)$ conjectured by Costin and coworkers~\cite{Costin2012} and mathematical approaches for $\epsilon_c=0$ by Basu and coworkers~\cite{Basu2014}. However, the systematic check-ups are missing and the setup of $\alpha=\beta=1$ is the worst choice to test the SB effect in the OBC since both boundaries generate the power-law tail in $\rho(x)$, so that there is no bulk even in the absence of the SB. In Ref.~\cite{Soh2017}, such finite-size effects were tested on $\rho(x)$ and $J$ with and without the SB for the PBC with $\rho_0=1/2$ and the OBC with $\alpha=\beta=1/2$. In particular, we focus on $J(\epsilon;L)$ with the comparison of exact solutions, $J(0;L)$\cite{Derrida1993,Derrida1993a,Blythe2007} and the analytic conjecture of $J(\epsilon)$\cite{Costin2012} in the thermodynamic limit.


In the intermediate SB regime, $0.2<\epsilon<0.5$, $\Delta_J$ is well fitted by the following analytic form, which is conjectured by Costin and coworkers~\cite{Costin2012} and confirmed in Ref.~\cite{Soh2017}:
\begin{equation}
\Delta_J\sim\exp(-b/\epsilon),
\label{eq:DJ-exp-fit}
\end{equation}
where $b\simeq 2$. However, we cannot say that this proves that $\epsilon_c=0$ because it does not fit the data well in the weak SB regime ($\epsilon<0.2$).  In order to resolve the finite-size effect in the SB problem, we employ two more numerical approaches in the conmbox of the KPZ universality in the following two subsections.  

\subsection{KPZ fluctuations with the SB}

We begin with the discussion of the SB effect in the transient regime of 1D BCSOS growth with $L$ sites as $\epsilon$ varies. The flat initial surface suffers surface relaxation until the system reaches the steady state at a finite time. During surface relaxation,  random deposition processes first lead the surface width to grow in time. As time elapses, the local interaction suppresses the width to grow slower than the random deposition case. 

\begin{figure}[]
	\centering
        \includegraphics[width=\textwidth]{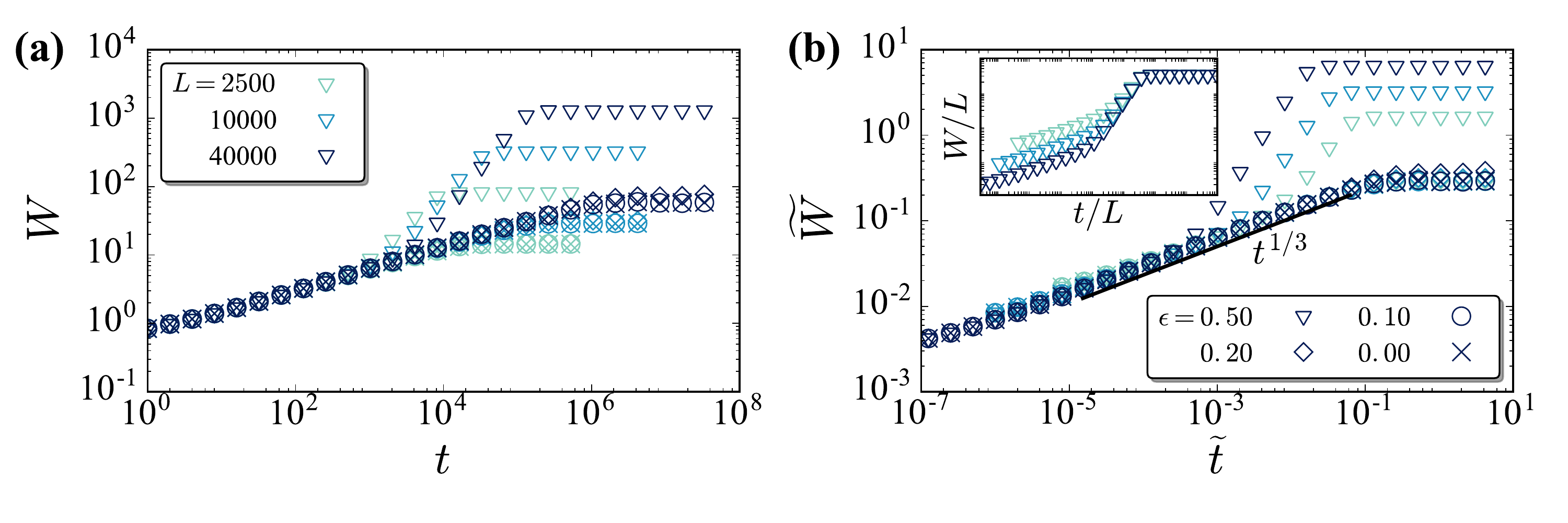}
	\caption{\label{fig:width} Dynamic scaling of surface width $W(\epsilon; L,t)$ is tested at $\epsilon = 0.50$ (triangle), $0.20$ (diamond), $0.10$ (circle) and $0$ (cross) for $L=2500\ \mathrm{to}\ 40000$ (from the lightest to the darkest):  (a) $W$ versus $t$ in double-logarithmic scales, and (b) the rescaled width $\widetilde{W} = WL^{-1/2}$ versus the rescaled time $\widetilde{t} = tL^{-3/2}$. The inset in (b) shows that another scaling collapse with $W/L$ versus $t/L$ works in the strong SB regime ($\epsilon=0.5$), which implies the global faceting in the steady state. All data are averaged over $10^8-10^{10}$ samples.}
\end{figure}

In the presence of the SB at the middle of the system, the height profile $h(x,t)$ forms a special shape near the SB, namely a facet, which spreads to the whole system and sharpen the surface width. Until the global facet forms, the surface grows as if there were no SB, however the density shock builds up and spreads away from the SB. This faceting sharpens the surface width in linear time as shown in Fig.~\ref{fig:width}. We here note that the time until the system grows a macroscopic facet is proportional to the system size, which is equivalent to the time scale that every site is influenced enough by the SB.

In the transient regime with sufficiently large time, the system fluctuates as the Gaussian orthogonal ensemble (GOE) Tracy-Widom distribution when the flat initial condition is employed. Once the SB is introduced in the system, the density difference around the SB velocitys up the surface to relax faster than sites far from the SB. The SB shock also drags the Tracy-Widom distribution to Gaussian one that is described by measuring skewness $S(x,t)$ and kurtosis $K(x,t)$ at each site at time $t$. 

\begin{figure}
	\centering
	\includegraphics[width=\textwidth]{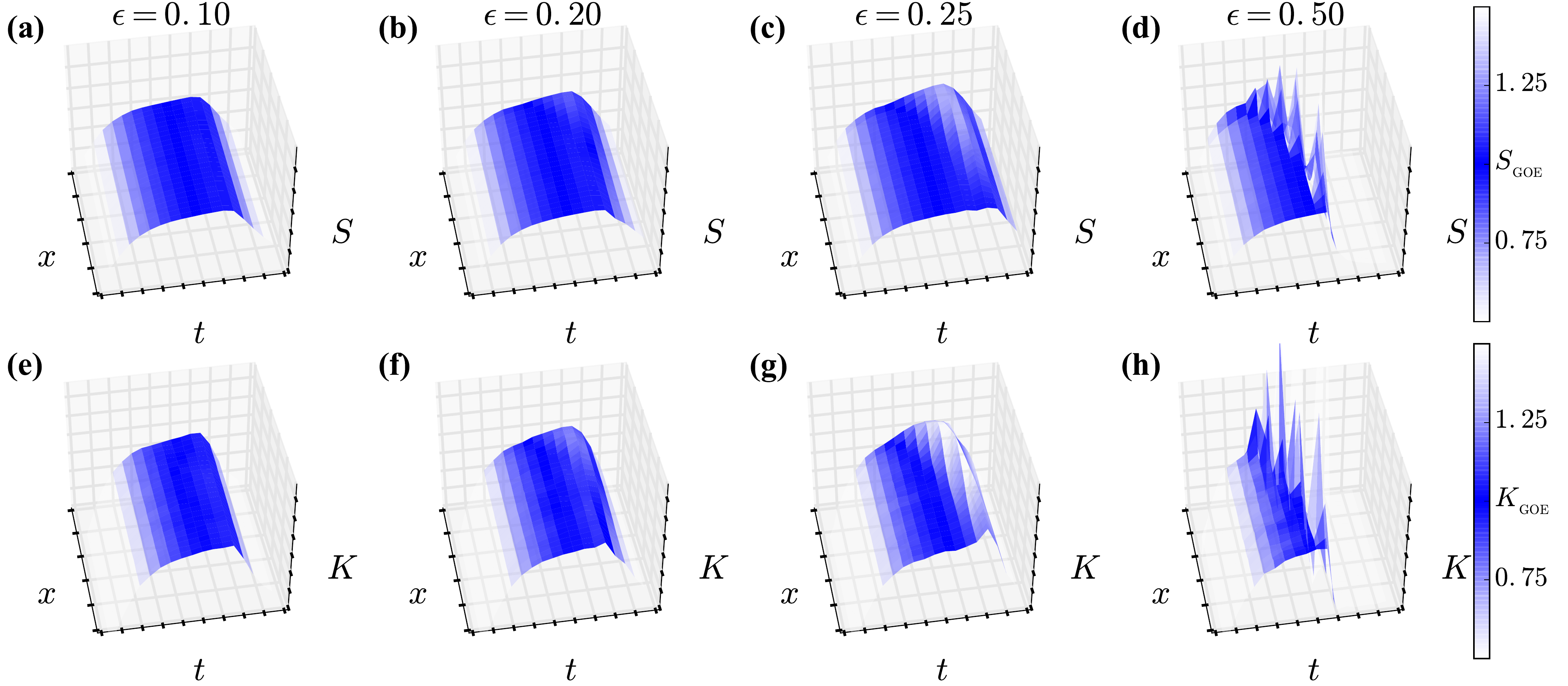}
        \caption{\label{fig:3D-SnK} Skewness $S(x,t)$ (a-d) and kurtosis $K(x,t)$ (e-f) of $h(x,t)$ for $L = 2500$ are plotted in the 3D format at $\epsilon = 0.10$ (a, e), $0.20$ (b, f), $0.25$ (c, g), and $0.50$ (d, h). Each spatial point is the value at every 50 sites away from the SB (from top to bottom) and every $2^n$ time steps (from left to right) from the flat initial condition. The color indicates ratio to $S_{_{\rm GOE}}$ and $K_{_{\rm GOE}}$ at the GOE Tracy-Widom distribution. Due to the finite-size effect, the system saturates to the KPZ fluctuations at $t < L^{3/2}$ and returns to the Gaussian fluctuations as the system reaches the stationary state. In the strong SB regime $(\epsilon=0.50\ \mathrm{and}\ 0.25)$, a shock propagates from the SB with the constant velocity as peaks move. Each simulation is averaged over $1.5 \times 10^6$ samples.}	
\end{figure}

\begin{figure}
	\centering
	\includegraphics[width=\textwidth]{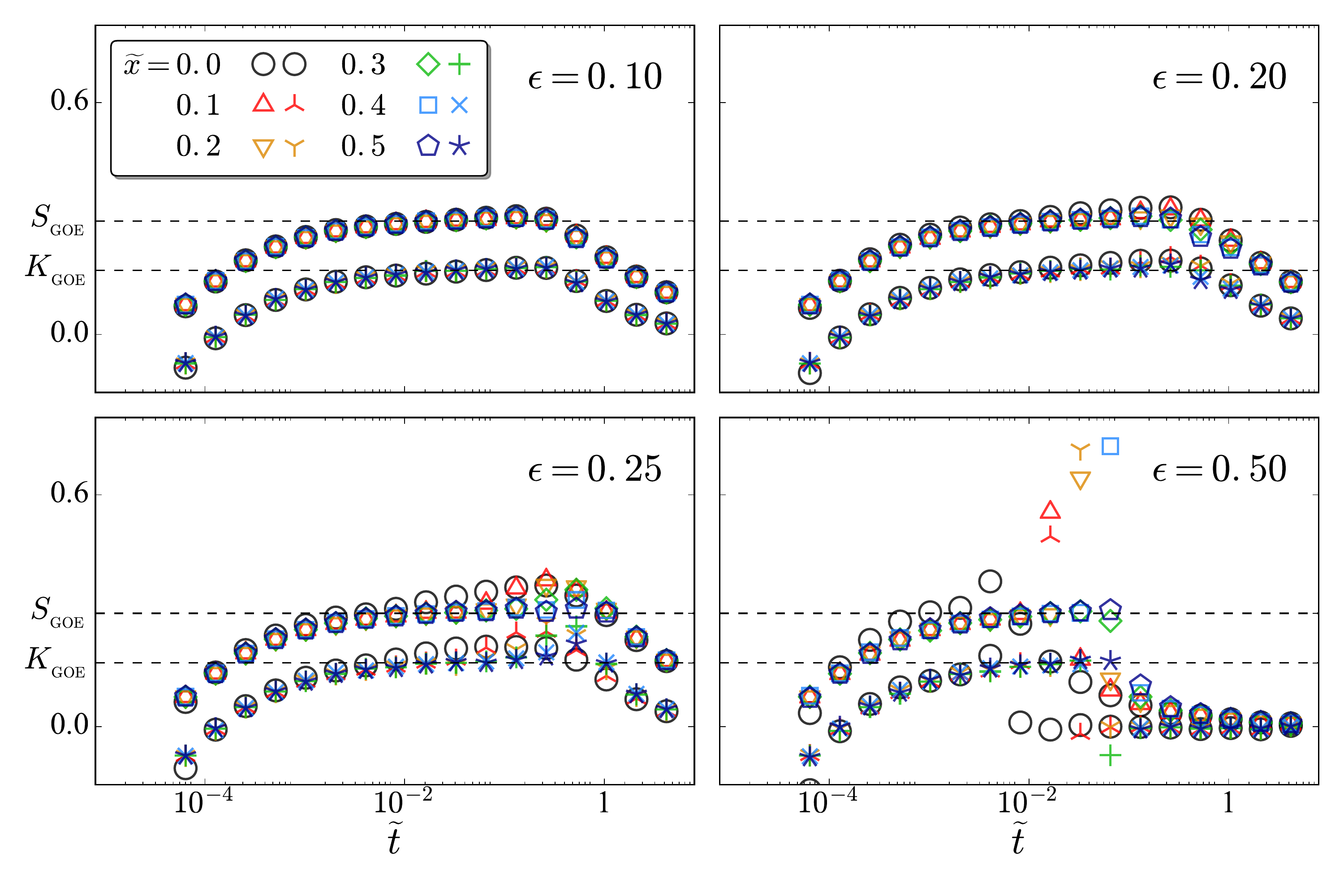}
        \caption{\label{fig:2D-SnK} $S(x,t)$ and $K(x,t)$ are plotted against $\tilde{t}=t/L^{3/2}$ for $L = 2500$ at $\epsilon = 0.10, 0.20, 0.25, \ \mathrm{and} \ 0.50$ for the selected values of $\tilde{x}=x/L$ from Fig.~\ref{fig:3D-SnK}.}	
\end{figure}

Figures~\ref{fig:3D-SnK} and \ref{fig:2D-SnK} show that the system can maintain the GOE values of $S_{_{\rm GOE}} \approx 0.2935$ and $K_{_{\rm GOE}} \approx 0.16524$  before the SB shock arrives. When the SB gets strong, the induced density field also becomes strong enough to relax faster than the global relaxation builds up, while the weak SB drags less, therefore, the system is relaxed almost uniformly.

%
\begin{figure}[]
	\centering 
	\includegraphics[width=0.75\textwidth]{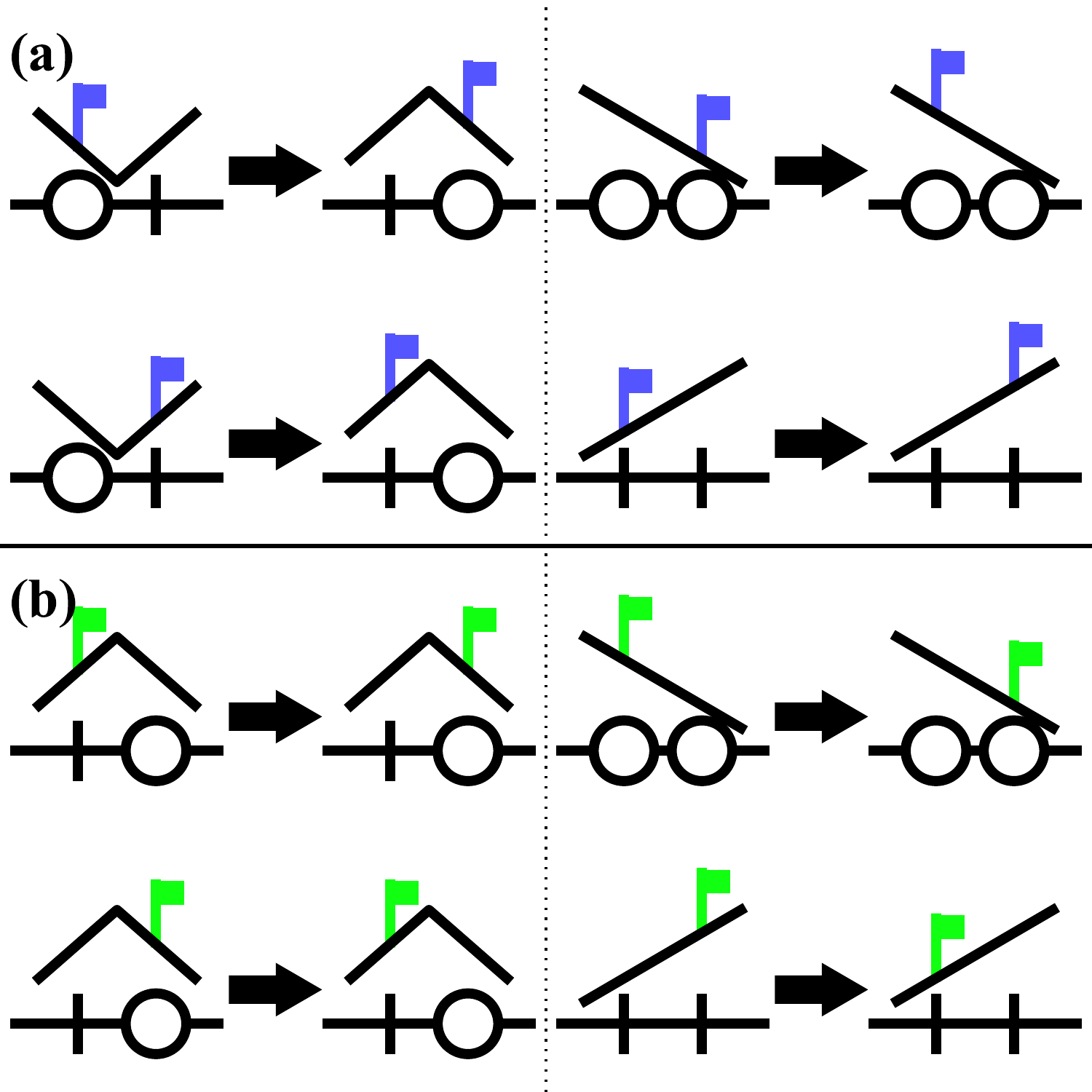}
	\caption{\label{fig:tracer} Dynamics of passive tracer: (a) positive feedback (advection) and (b) negative feedback (anti-advection). We here illustrate all the possible movements of the tracer and the the blue/green flag represents the positive/negative tracer.}
\end{figure}
%
\begin{figure}[]
	\centering
	\includegraphics[width=0.8\textwidth]{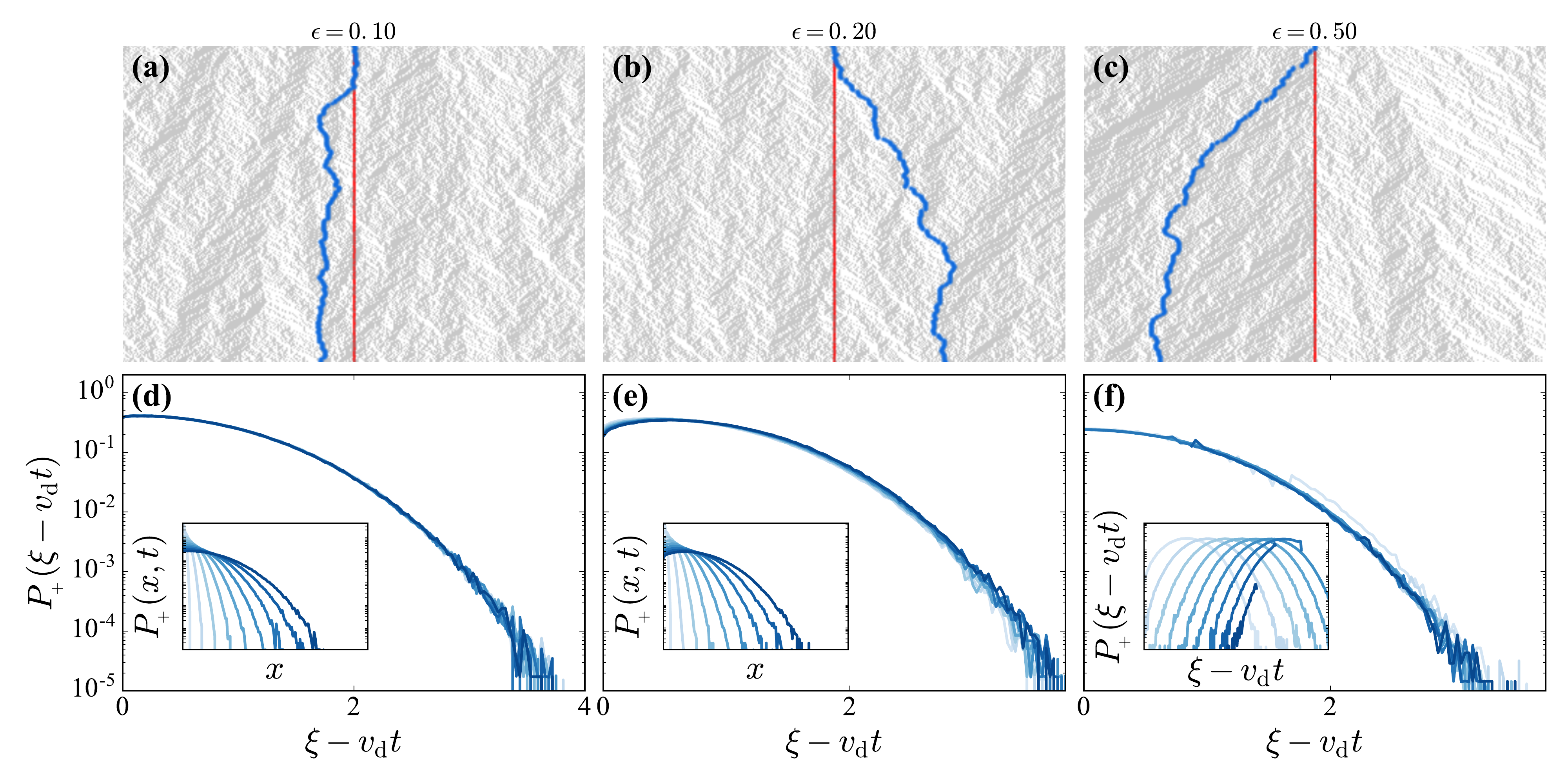}
	\caption{\label{fig:tracer1} In the presence of particles as fields, at the SB strength $\epsilon = 0.10 \ \mathrm{(weak)},\ 0.20\ \mathrm{(intermediate)},\ \mathrm{and}\ 0.50\ \mathrm{(strong)}$ (a-c), we present spatiotemporal movements of the positive tracer, where each dot and space in horizontal lines represent site configurations with the SB indicated as the red strip at the middle. An instance of the positive tracer is represented as the blue trail.  (d-f) The rescaled displacement $\xi = xt^{-1/z_p}$ distribution are presented and the insets show the original distribution of $P_+(x,t)$. Both of the panels are drawn on the half plane. Each distribution is averaged over $10^6$ samples.}
\end{figure}
\begin{figure}[]
	\centering
	\includegraphics[width=0.9\textwidth]{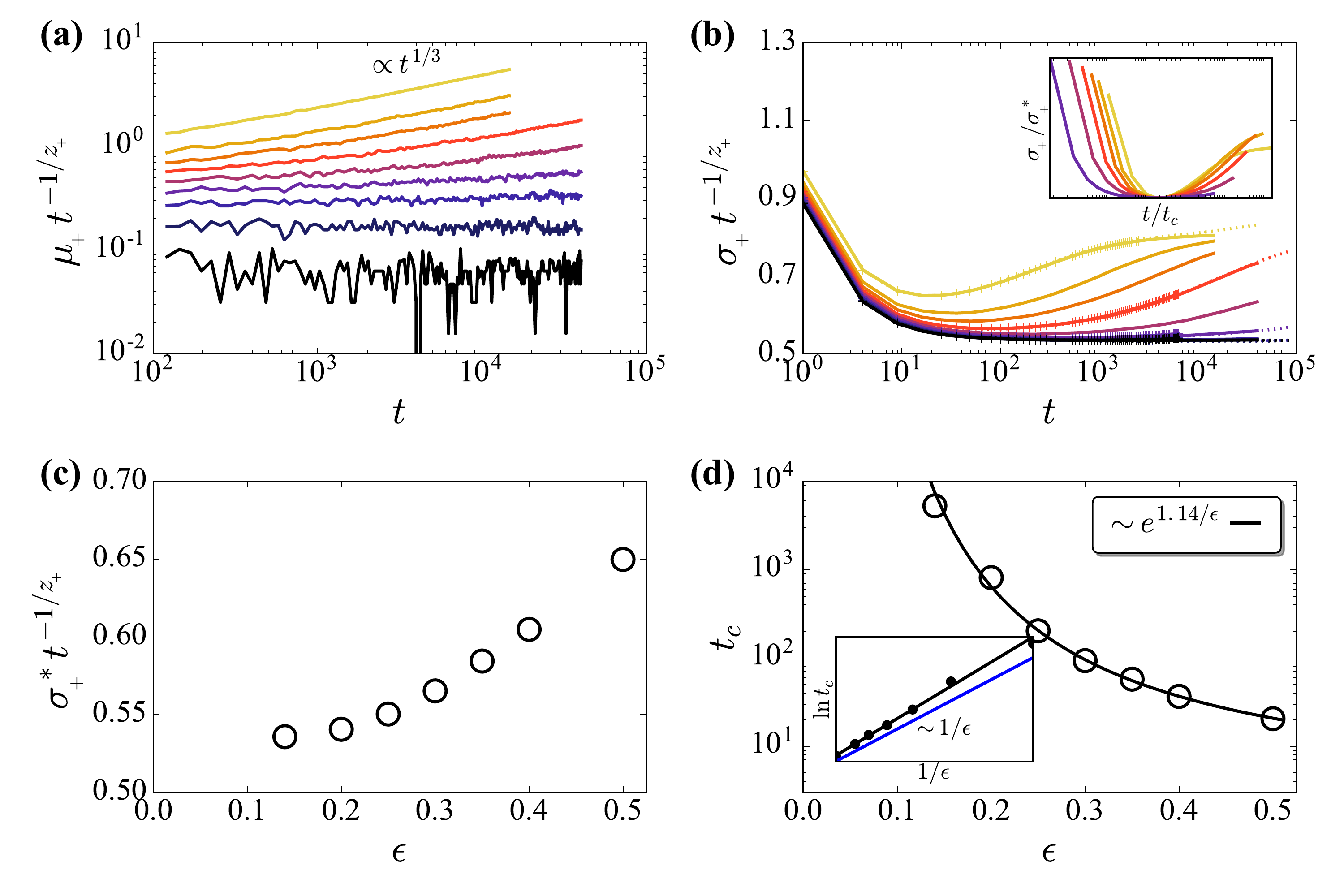}
	\caption{\label{fig:tracer1+} The statistics analyses of Fig.~\ref{fig:tracer1} for the positive tracer are presented at $\epsilon = 0, 0.1, 0.14, 0.2, 0.25, 0.3, 0.35, 0.4,  \mathrm{and}\ 0.5$ (from the darkest/bottom to the lightest/top) in the OBC for $L=10000$: (a) $\mu_p t^{-1/z_p}$ versus $t$ in double-logarithmic scales. In the strong SB regime (bright lines) evolves as $t^{1/3}$, which implies a constant drift from the SB. (b) $\sigma_p t^{-1/z_p}$ versus $t$ in the semi-logarithmic scale. The time at the minimal value of each curve is indicated as the crossover time $t_c$. In the inset of (b), $\sigma_p/\sigma^*_p$ versus $t/t_c$. (c) $\sigma_p^*$ versus $\epsilon$ and (d) $t_c$ versus $\epsilon$, where grey shaded dots represent the limiting data values (too erroneous to estimate). The inset of (d) shows the least squares fitting (LSF) of 
$\ln(t_c)$ with $1/\epsilon$, by coefficient $1.144$, where the blue line guides the slope of $1$.}
\end{figure}
\begin{figure}
	\centering
	\includegraphics[width=0.8\textwidth]{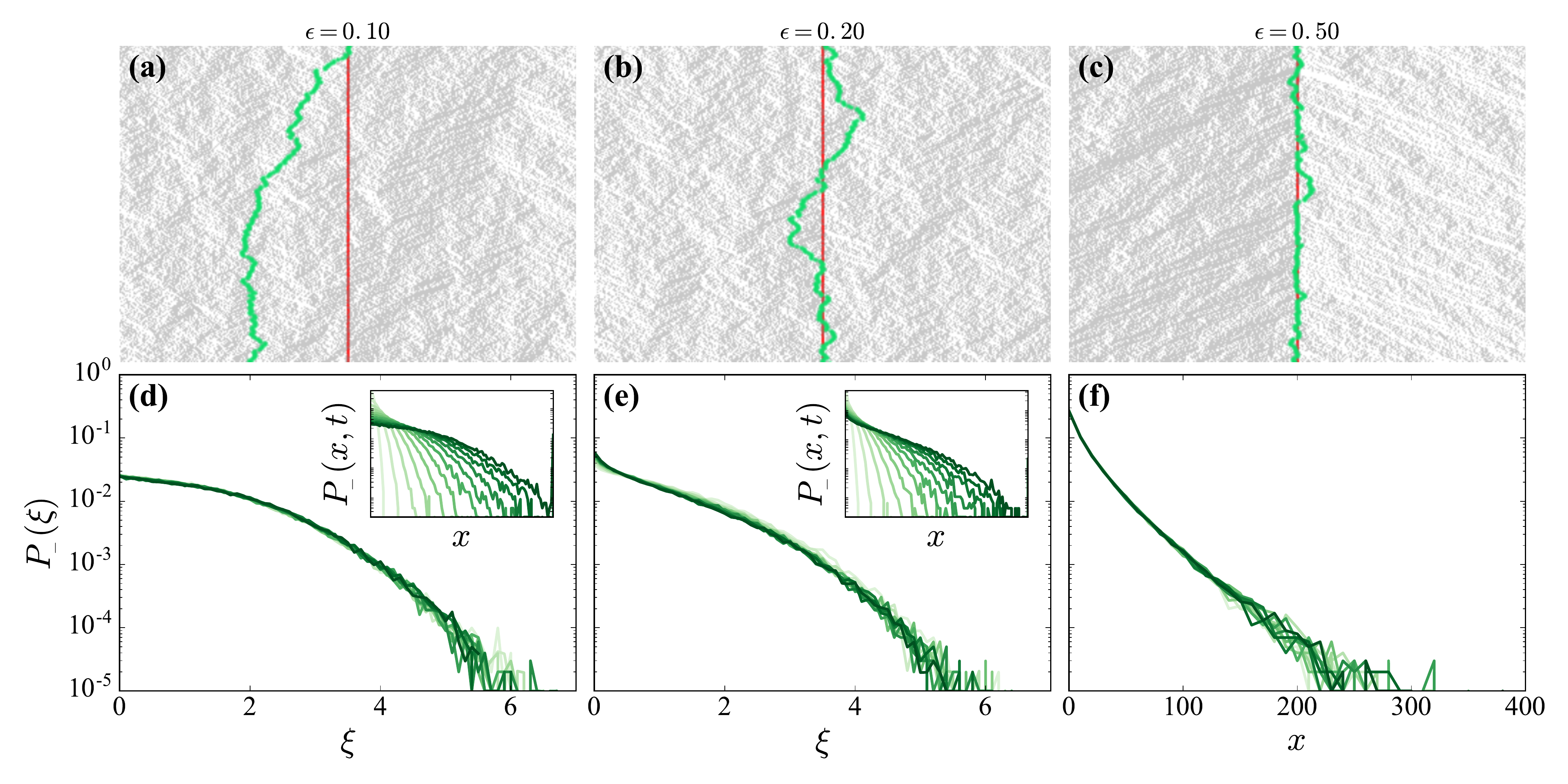}
	\caption{\label{fig:tracer2} In the presence of particles as fields, at the SB strength $\epsilon = 0.10 \ \mathrm{(weak)},\ 0.20\ \mathrm{(intermediate)},\ \mathrm{and}\ 0.50\ \mathrm{(strong)}$ (a-c), we present spatio-temporal movements of the negative tracer, where each dot and space in horizontal lines represent site configurations with the SB indicated as the red strip at the middle. An instance of the negative tracer is represented as the green trail.  (d-f) The rescaled displacement ($\xi = xt^{-1/z_m}$) distribution are presented and the insets show the original distribution of $P_-(x,t)$. Note that the negative tracer does behave as neither particle nor hole. Each distribution is averaged over $5 \times 10^4$ samples.}
\end{figure}
\begin{figure}
	\centering
	\includegraphics[width=0.9\textwidth]{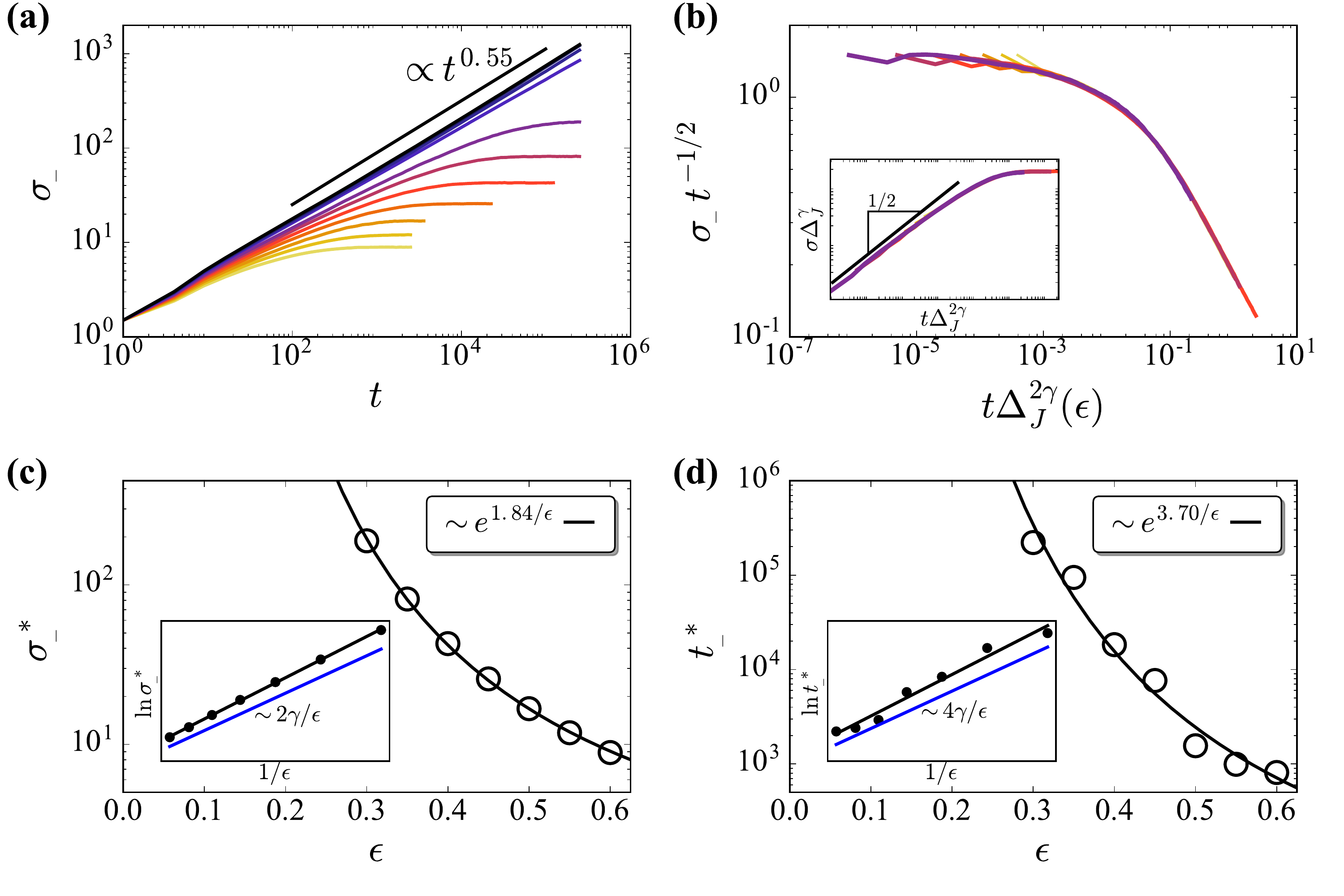}
	\caption{\label{fig:tracer2-} The statistics analyses of Fig.~\ref{fig:tracer2} for the negative tracer are presented at $\epsilon = 0, 0.1, 0.14, 0.2, 0.3, 0.35, 0.4, 0.45, 0.5, 0.55,$ and 0.6 (from the darkest to the lightest) in the OBC for $L=10000$. (a) $\sigma_p(t)$ versus $t$ in double-logarithmic scales. In the weak SB regime, the tracer anomalous diffuses with $1/z_m \approx 0.55$, while in the strong SB regime ($\epsilon>0.2$), (b) data collapse works well in terms of the rescaled standard deviation $\sigma_m t^{-1/2}$ and the rescaled time $t\Delta^2_J$. The inset shows that the slope of data collapse is $\sim 0.5$, which denotes that the saturation is closely related to the standard diffusion, $1/z=0.5$. (c-d) The minimal value of the saturated standard deviation $\sigma^*_m$ and the saturation time $t^*_m$ are plotted against the SB strength $\epsilon$, respectively. The data in the insets of (c-d) obtained from (b) are fitted by the LSF of $\ln(\sigma^*_m)\sim\left[\ln(t^*_m)\right]^{1/2}\sim b/\epsilon$ with $b\approx 2$: $1.837 \mbox{ for (c) and }3.699 \mbox{ for (d)}$, where blue lines guide the slope of $2 \mbox{ and } 4$, respectively.}
\end{figure}
\subsection{Tracer dynamics with the SB}

In this subsection, we employ a passive tracer to figure out the SB effect in detail. In particular, we consider two types of passive tracers: one is with the positive feedback from surface growth, which slides with the avalanche of the growth (advection in fluid dynamics), and the other is with the negative feedback from surface growth, which slides against the avalanche (anti-advection).  

Similar to earlier studies\cite{Drossel2002,CSChin2002,HKim2011} for the dynamics of passive tracers in 1D KPZ interface without the SB, we define a positive (negative) tracer in the TASEP (1D BCSOS growth) with the SB as illustrated in Fig.~\ref{fig:tracer} (a). Due to the SB, the underlying TASEP suffers jamming, which induces the bias of the field around the SB. Two  following questions are posed as the core of the SB problem: ``Is $\epsilon_c$ finite?" and ``How the SB change the nature of the KPZ universal scaling in tracer dynamics?"

\subsubsection*{\bf Positive Tracer}
--
As denoted in Eq.~(\ref{eq:tracer}), the dynamics of the positive tracer, 
see Fig.~\ref{fig:tracer} (a) is equivalent to that of the second-class particle in the TASEP~\cite{Derrida1993a}. Note that this tracer is always on either particle (upper cases) or hole (lower cases). For the case of the tracer dynamics defined in Ref.~\cite{Drossel2002}, the particle on a surface should select one of the neighbor site if both NN site has the same height, see the left cases of Fig.~\ref{fig:tracer} ), where in this case random selection is restricted.  The passive tracer is super-diffusive, so the standard deviation $\sigma_p(t)=(\Delta x)\sim t^{1/z_p}$ with the dynamic exponent $z_p = z_{_{\rm KPZ}}=3/2$. As the SB gets strong, the positive tracer experiences repulsive forces from the SB, resulting a constant drift from the SB. In Fig.~\ref{fig:tracer1}, we present the spatio-temporal movements for a typical positive tracer and the distribution of its displacement at time $t$, $P_+(x,t)$. The displacement can be rescaled as $\xi =xt^{-1/z_p}$. Therefore, the distribution of $P_+(x,t)$ collapses into the single curve with the constant drift, $\xi'=\xi-v_g$, that depends on $\epsilon$ as shown in Fig.~\ref{fig:tracer1}.

In Fig.~\ref{fig:tracer1+}, we analyze the statistical properties of $P_+(x,t)$ systematically to discuss how the impact of the SB spreads to the whole system. Here we distinguish that a constant drift of the positive tracer is apparent in the strong SB regime $\epsilon > 0.2$, while below the some point of the SB strength $\epsilon$, say $\epsilon<0.2$, the tracer seems to have a very slow drift, or even none. The standard deviation of the positive tracer position at time $t$, $\sigma_p(t)$, without the SB effect, scales as $\sigma_p(t)\sim t^{2/3}$ with the KPZ dynamic exponent $z=3/2$, while the strong SB effect makes the positive tracer to lead the ballistic diffusion, so $\sigma_p(t)\sim t$.

Since the positive tracer does not reside the outside of a particle (as well as a vacancy, by the particle-vacancy symmetry), the diffusion characteristics remains that of the underlying TASEP field. In the biased density field, it does drift away from the SB, so that the displacement distribution cannot become stationary. Moreover, the minimum of $\sigma_p(t)$ indicates the time of particles drifting out around the SB. Again, we note that this crossover time $t_c$ is proportional to inverse of $v_g$: 
\begin{equation}
t_c \sim v_g^{-1}(=\Delta_b^{-1}=\Delta_J^{-1/2}) \sim \exp(1/\epsilon).
\end{equation} 

\subsubsection*{\bf Negative Tracer}
--
The same measurements are analyzed for the negative tracer, the dynamics of which is illustrated in Fig.~\ref{fig:tracer}~(b). Negative tracer moves as opposite orientation to positive tracer in case of surface neighboring surface has different height. However when the neighboring height is the same particle moves as given rule, because there is no time-reversal picture of underlying particle dynamics. As $\epsilon$ increases, fluctuations becomes suppressed by the biased density field around the SB to the dynamic exponent $z_m >z_{_{\rm KPZ}}=3/2$. This behavior is distinct from that of the positive tracer. The negative tracer does not suffer any constant drift due to the convergent field, therefore, it is not able to move ballistically.

As shown in Fig.~\ref{fig:tracer2}, the typical spatio-temporal movements of the negative tracer is confined in a certain domain when the SB strength is strong enough. This behavior is analyzed by the statistics of the displacement distribution for the negative tracer at time $t$, such as the standard deviation $\sigma_m(t)$. In the strong SB regime where macroscopic jamming is apparent, we observe that tracer is confined within a finite time as a random walker in the confined space, while in the weak SB regime, it anomalously diffuses even in the presence of the SB. In Fig.~\ref{fig:tracer2-}, we present that $\sigma_m(t)$ scales as $\sigma_m(t)\sim t^{1/z_m}$. Without the SB, the diffusion is characterized by $z_m \simeq 1.8$, which is within the range reported by previous work $(z=1.74 \sim 1.98)$~\cite{Drossel2002}.  

Moreover, uncorrelated approximations are proposed to explain the asymptotic displacement distribution for the negative tracer in the SB biased TASEP Field. The negative tracer has the case of changing the orientation from a particle to a vacancy and vice versa. The negative tracer driven 
by the SB biased particle field~\cite{Drossel2002} is described in Eq.~(\ref{eq:tracer}), 
where the driving field includes its own intrinsic randomness. Hence, we apply uncorrelated approximations to the field as follows:
\begin{equation}
\frac{dx(t)}{dt} = -\left[v(x;\epsilon) \right]_{x}+ \zeta(t).
\end{equation}
The dynamics of the tracer can be written by a Fokker-Planck (FP) equation as follows:
\begin{equation}
\frac{\partial P_-(x,t)}{\partial t} = K\frac{\partial^2 P_-(x,t)}{\partial x^2} + v_g\mbox{sign}(x)\frac{\partial P_-(x,t)}{\partial x},
\end{equation}
where $K$ is a diffusion constant. The group velocity $v_g$ is equivalent to the global slope in the BCSOS growth.

Regardless of the initial distribution, the stationary solution sis as follows:
\begin{equation}
P_-(x) = \frac{v_g}{2}\exp(-v_g|x|), 
\end{equation}
where $v_g=\Delta_b=\Delta_J^{1/2}$. The MF value for the standard deviation of this distribution is 
$$\sigma_{_{\rm MF}}=\sqrt{2}/v_g \sim \Delta_b^{-1}\sim \Delta_J^{-1/2}.$$ 
Using the  the MF current $J_{_{\rm MF}}=\rho_b(1-\rho_b)$ and the approximate conjecture for $\Delta_J\sim\exp(b/\epsilon)$~\cite{Costin2012} with $b\approx 2$, we are able to derive 
\begin{equation}
\sigma_{_{\rm MF}} = C\exp(1/\epsilon),
\end{equation}
where $C$ is a constant. 

\subsubsection*{\bf Fractional FP equation with intuitive arguments}
--
However, the MF result is slightly different from our numerical observation. It is because we ignore the anomalous diffusion nature of the tracer. To take account of super-diffusive behavior, we establish a fractional FP (fFP) equation,
\begin{equation}
\frac{\partial P_-(x,t)}{\partial t} = \frac{\partial}{\partial x}\left[KD^{1-2/z_m}_{t}\frac{\partial}{\partial x} +v_g\mbox{sign}(x)\right]P_-(x,t),
\label{eq:fFP}
\end{equation}
where $K$ is a diffusion constant and the operator $D^{1-2/z_m}_{t}$ is the Riemann-Liouville fractional differential operator of order $(1-2/z_m)$ about $t$. 

\begin{figure}[]
	\centering
	\includegraphics[width=0.75\textwidth]{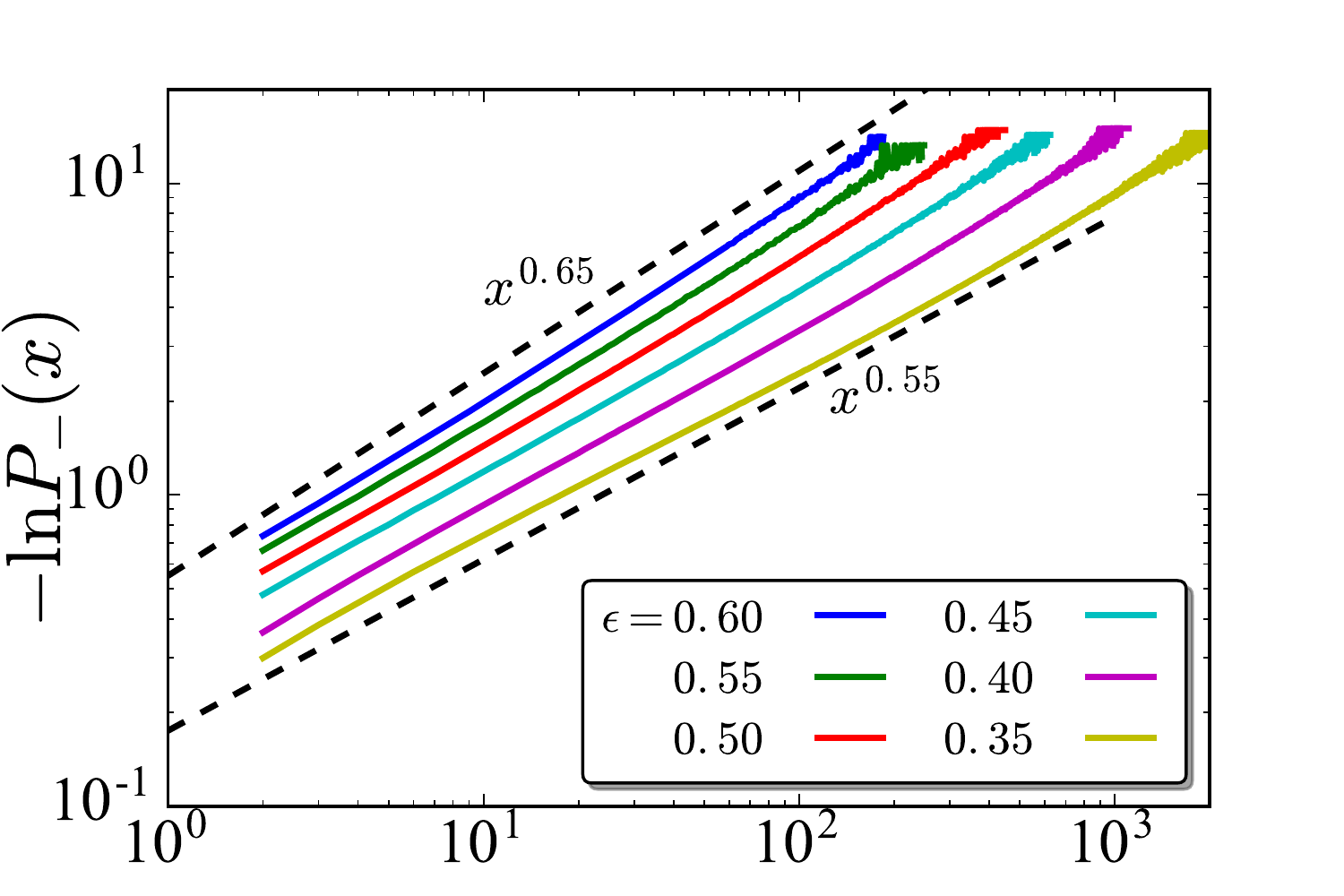}
	\caption{\label{fig:fFP} The stationary stretched exponential distribution function $P_-(x)$ for the negative tracer are plotted as $\ln P_-(x)$ versus $x$ in double-logarithmic scales, where $\gamma\in[0.55,0.65]$ against $\epsilon\in [0,65,0.35]$.}
\end{figure}

The fFP equation, Eq.~(\ref{eq:fFP}), gives the solution $P_-(x,t)$ with the following statistical properties:
\begin{eqnarray}
\mu_m(t)=\left \langle x \right \rangle& \sim& t, \\
\sigma^2_m(t)=\left \langle (\Delta x)^2 \right \rangle& \sim& t^{2/z_m},
\end{eqnarray}
which represents a constant drift (mean displacement) and an anomalous diffusion (fluctuation) of $P_-(x,t)$. In the spirit of this anomalous diffusion, the time-differential operator is related to the spatial derivative via $dt \sim dx^{z_m}$, so 
the first term becomes the fractional derivative about space via 
\begin{eqnarray}
\left( \frac{\partial}{\partial t}\right)^{1-\frac{2}{z_m}}\left(\frac{\partial}{\partial x}\right) &=& \left( \frac{\partial}{\partial x}\right)^{z_m(1-\frac{2}{z_m})}\left( \frac{\partial}{\partial x}\right)\nonumber \\
 &=& \left( \frac{\partial}{\partial x} \right)^{z_m-1}.
\end{eqnarray}
In the steady-state limit, Eq.~(\ref{eq:fFP}) satisfies 
\begin{equation}
\label{eq:SteadyEq}
0 = \frac{\partial}{\partial x}\left[K\left(\frac{\partial}{\partial x}\right)^{\gamma}+v_g\mbox{sign}(x) \right] P_-(x).
\end{equation}
Note that although the first derivative leads $P_-(x)$ to have an arbitrary constant, the constant should always be 0 due to normalization of the probability. Therefore, the resulting $P_-(x)$ has the stretched exponential form as follows:
\begin{equation}
P_-(x) = C_-\exp(-\Delta_b|x|^\gamma)
\end{equation}
where $\gamma = z_m-1$ and $C_-$ is a proper normalization constant. As a result, we can find that the stationary standard deviation $\sigma^*_m$ that scales as 
\begin{equation}
\sigma^*_m = \sqrt{\frac{\Gamma(3/\gamma)}{\beta \Gamma(1+1/\gamma)}}\Delta_b^{-1/\gamma}\sim \Delta_J^{-1/2\gamma}.
\end{equation} 
This relation connects the stretched exponential distribution function $P_-(x)$ with $\gamma \simeq 0.6$, see Fig.~\ref{fig:fFP} to $b\simeq 2.21$ in $\sigma_m^*\sim \exp(b/2\gamma\epsilon)$ based on $\Delta_J\sim \exp(-b/\epsilon)$. This corresponds to $z_m \simeq 1.6$, which is still in the intermediate stage between random (normal) diffusion and KPZ, $z_{_{\rm KPZ}}(=3/2) < z_m < 2 ~\mbox{(normal diffusion)}$.

Another possible origin to the stretched-exponential distribution is the inhomogeneity in the density profile near the SB. In this case, we assume again that particle movements fluctuate by random Gaussian noise, but the drift term $v_g$ is actually not constant. From Eq.~(\ref{eq:DP}), $v_g=\Delta(x)$, so that Eq.~(\ref{eq:SteadyEq}) can be rewritten as
\begin{equation}
0 = \frac{\partial}{\partial x}\left[K\left(\frac{\partial}{\partial x}\right)+v_g(|x|)\mbox{sign}(x) \right] P_-(x).
\label{eq:FP+DP}
\end{equation}

Recalling Eq.~(\ref{eq:DP}), $\Delta(x) = \Delta_b + Ax^{-\nu}$, with the assumption $P_-(x) \propto \exp(-\int\Delta(x)dx)$, the solution of Eq.~(\ref{eq:FP+DP}) reduces the following simple form:
\begin{equation}
P_-(x) = C'_- \exp \left[-\Delta_b |x|-A/(1-\nu)|x|^{(1-\nu)}\right],
\end{equation}
where $C_-'$ is a proper normalization constant. This also agrees well with the value of $\nu$ in both intermediate and strong SB regimes, $\nu\in[0.33, 0.5]$ as $\Delta_b\to 0$ with $\epsilon\to\epsilon_c$. Our numerical results are presented in Fig.~\ref{fig:fFP}. Morever, we provide the summary of the comparison of numerical results with analytic conjectured values as Table~\ref{tbl:conjecture} for various observables.
\begin{table}[]
\centering
\caption{\label{tbl:conjecture} The comparison of analytical conjecture~\cite{Costin2012} and the LSF values of numerical simulations in the strong SB regime. Here $\gamma=z_m-1\simeq 0.6$ for the negative tracer as shown in Fig.~\ref{fig:fFP}. }
\begin{tabular}{cccc}
\hline\hline
Variable & Conjecture & b (LSF) & ~~Ratio\\
\hline\hline
$\Delta_J=[\rho_b(1-\rho_b)-\frac{1}{4}]$ & $\sim\exp(-b/\epsilon)$ & $b\simeq 2.25$ & 1 \\  
$t_c\sim v_g^{-1}\sim \Delta_J^{-1/2}$ & $\sim \exp(b/2\epsilon)$ & $b\simeq 2.29$ & $1.02$ \\ 
$\sigma^*_m\sim \Delta_J^{-1/2\gamma}$ & $\sim \exp(b/2\gamma\epsilon)$ & $b\simeq 2.21 $ & $ 0.98$ \\ 
$^{\rm a}$ $ t^*_m\sim (\sigma^*_m)^2 \sim \Delta_J^{-2}$ & $\sim \exp(b/\gamma\epsilon)$ & $b\simeq 2.27 $ & $ 1.01 $ \\ 
\hline\hline
\end{tabular}
\end{table}
\section{Summary with Remarks}
\label{sec:conclusion}

To sum up, we revisited so far the slow-bond (SB) problem in the totally asymmetric simple exclusion process on a one-dimensional lattice. We considered it with both open and closed boundary conditions, in the context of the existence of the non-queued SB phase in the thermodynamic limit. This SB problem, has a longstanding controversial issues exist because of some discrepancy between numerical results and mathematical results with some approximations. In particular, we discussed the SB relevance from fundamental relations, \textit{i.e.}, density and current, with derivative observables (positive/negative tracers) to observe how the Kardar-Parisi-Zhang (KPZ) universality properties are deformed. Our extensive numerics showed detailed structures related to the finite-size SB effect, and checked out the ensemble equivalency as well. The anomalous diffusion of the tracer can be proposed another comprehensive view of the SB problem. Extensions of this problem, such as complex defects or controlling junctions is an interesting subject for further studies. As the final remark, we would like to comment on something differences happened near $\epsilon\approx 0.2$ as if there is either a phase transition or a crossover between weak and strong SB effects. To numerically clarify the issue is currently not doable. However, we found that the ambiguity of the drift velocity plays a crucial role in the SB problem. The proper choice of the drift velocity determines a boundary for either a phase transition or a crossover occurs. 
%

%
\ack{This research was supported by Basic Science Research Program through the National Research Foundation of Korea (NRF)~(KR) [NRF-2017R1D1A3A03000578 (M.H.)] and contained unpublished results of the first author's dissertation. We would also like to acknowledge valuable comments from T. Sasamoto, J.M. Kim, and Y. Baek.}

\bibliography{iopart-num}

\providecommand{\newblock}{}
\begin{thebibliography}{10}
\expandafter\ifx\csname url\endcsname\relax
  \def\url#1{{\tt #1}}\fi
\expandafter\ifx\csname urlprefix\endcsname\relax\def\urlprefix{URL }\fi
\providecommand{\eprint}[2][]{\url{#2}}

\bibitem{Kardar1986}
Kardar M, Parisi G and Zhang Y~C 1986 {\em Phys. Rev. Lett.\/} {\bf 56}
  889--892 \urlprefix\url{http://link.aps.org/doi/10.1103/PhysRevLett.56.889}

\bibitem{Meakin1986}
Meakin P, Ramanlal P, Sander L~M and Ball R~C 1986 {\em Phys. Rev. A\/} {\bf
  34} 5091--5103
  \urlprefix\url{http://link.aps.org/doi/10.1103/PhysRevA.34.5091}

\bibitem{Krug2000}
Krug J 2000 {\em Braz. J. Phys.\/} {\bf 30} 97--104
  \urlprefix\url{http://www.scielo.br/scielo.php?script=sci_arttext&pid=S0103-97332000000100009&nrm=iso}

\bibitem{Kardar1987}
Kardar M and Zhang Y~C 1987 {\em Phys. Rev. Lett.\/} {\bf 58}(20) 2087--2090
  \urlprefix\url{http://link.aps.org/doi/10.1103/PhysRevLett.58.2087}

\bibitem{Johansson2000}
Johansson K 2000 {\em Communications in Mathematical Physics\/} {\bf 209}
  437--476 ISSN 1432-0916
  \urlprefix\url{http://dx.doi.org/10.1007/s002200050027}

\bibitem{Bodineau2005}
Bodineau T and Martin J 2005 {\em Electron. Commun. Probab.\/} {\bf 10}
  105--112 \urlprefix\url{http://dx.doi.org/10.1214/ECP.v10-1139}

\bibitem{Takeuchi2010}
Takeuchi K~A and Sano M 2010 {\em Phys. Rev. Lett.\/} {\bf 104}(23) 230601
  \urlprefix\url{http://link.aps.org/doi/10.1103/PhysRevLett.104.230601}

\bibitem{Takeuchi2011}
Takeuchi K~A, Sano M, Sasamoto T and Spohn H 2011 {\em Scientific Reports\/}
  {\bf 1} 34 article \urlprefix\url{http://dx.doi.org/10.1038/srep00034}

\bibitem{Tang1993}
Tang L~H and Lyuksyutov I~F 1993 {\em Phys. Rev. Lett.\/} {\bf 71} 2745--2748
  \urlprefix\url{http://link.aps.org/doi/10.1103/PhysRevLett.71.2745}

\bibitem{Janowsky1994}
Janowsky S~A and Lebowitz J~L 1994 {\em J. Stat. Phys.\/} {\bf 77} 35--51
  \urlprefix\url{http://dx.doi.org/10.1007/BF02186831}

\bibitem{MHa2003}
Ha M, Timonen J and den Nijs M 2003 {\em Phys. Rev. E\/} {\bf 68} 056122
  \urlprefix\url{http://link.aps.org/doi/10.1103/PhysRevE.68.056122}

\bibitem{Costin2012}
Costin O, Lebowitz J~L, Speer E~R and Troiani A 2013 {\em Bull. Inst. Math.,
  Acad. Sin. (New Series)\/} {\bf 8} 49
  \urlprefix\url{https://w3.math.sinica.edu.tw/bulletin_ns/20131/2013103.pdf}

\bibitem{Basu2014}
{Basu} R, {Sidoravicius} V and {Sly} A 2014 {\em ArXiv e-prints\/}
  (\textit{Preprint} \eprint{1408.3464})

\bibitem{Schmidt2015}
Schmidt J, Popkov V and Schadschneider A 2015 {\em EPL (Europhysics Letters)\/}
  {\bf 110} 20008
  \urlprefix\url{http://stacks.iop.org/0295-5075/110/i=2/a=20008}

\bibitem{Soh2017}
Soh H, Baek Y, Ha M and Jeong H 2017 {\em Phys. Rev. E\/} {\bf 95}(4) 042123
  \urlprefix\url{https://link.aps.org/doi/10.1103/PhysRevE.95.042123}

\bibitem{Parmeggiani2004}
Parmeggiani A, Franosch T and Frey E 2004 {\em Phys. Rev. E\/} {\bf 70} 046101
  \urlprefix\url{http://link.aps.org/doi/10.1103/PhysRevE.70.046101}

\bibitem{Embley2009}
Embley B, Parmeggiani A and Kern N 2009 {\em Phys. Rev. E\/} {\bf 80} 041128
  \urlprefix\url{http://link.aps.org/doi/10.1103/PhysRevE.80.041128}

\bibitem{Brankov2004}
Brankov J~G, Pesheva N~C and Bunzarova N 2004 {\em Phys. Rev. E\/} {\bf 69}
  066128 \urlprefix\url{http://link.aps.org/doi/10.1103/PhysRevE.69.066128}

\bibitem{Neri2011}
Neri I, Kern N and Parmeggiani A 2011 {\em Phys. Rev. Lett.\/} {\bf 107} 068702
  \urlprefix\url{http://link.aps.org/doi/10.1103/PhysRevLett.107.068702}

\bibitem{Basu2010}
Basu M and Mohanty P~K 2010 {\em J. Stat. Mech.\/} {\bf 2010} P10014
  \urlprefix\url{http://stacks.iop.org/1742-5468/2010/i=10/a=P10014}

\bibitem{Balents1994}
Balents L and Kardar M 1994 {\em Phys. Rev. B\/} {\bf 49} 13030--13048
  \urlprefix\url{http://link.aps.org/doi/10.1103/PhysRevB.49.13030}

\bibitem{Foulaadvand2008}
Foulaadvand M~E, Kolomeisky A~B and Teymouri H 2008 {\em Phys. Rev. E\/} {\bf
  78} 061116 \urlprefix\url{http://link.aps.org/doi/10.1103/PhysRevE.78.061116}

\bibitem{Kinzelbach1995}
Kinzelbach H and L{\"a}ssig M 1995 {\em J. Phys. A: Math. Gen.\/} {\bf 28} 6535
  \urlprefix\url{http://stacks.iop.org/0305-4470/28/i=23/a=009}

\bibitem{Hwa1995}
Hwa T and Nattermann T 1995 {\em Phys. Rev. B\/} {\bf 51} 455--469
  \urlprefix\url{http://link.aps.org/doi/10.1103/PhysRevB.51.455}

\bibitem{Lassig1998}
L{\"a}ssig M 1998 {\em J. Phys.: Cond. Matter\/} {\bf 10} 9905
  \urlprefix\url{http://stacks.iop.org/0953-8984/10/i=44/a=003}

\bibitem{JHLee2009}
Lee J~H and Kim J~M 2009 {\em Phys. Rev. E\/} {\bf 79} 051127
  \urlprefix\url{http://link.aps.org/doi/10.1103/PhysRevE.79.051127}

\bibitem{Seppalainen2001}
Sepp{\"a}l{\"a}inen T 2001 {\em Journal of Statistical Physics\/} {\bf 102}
  69--96 ISSN 1572-9613
  \urlprefix\url{http://dx.doi.org/10.1023/A:1026508625058}

\bibitem{Kandel1992}
Kandel D and Mukamel D 1992 {\em Europhysics Letters\/} {\bf 20} 325--329
  \urlprefix\url{http://gateway.webofknowledge.com/gateway/Gateway.cgi?GWVersion=2&SrcAuth=mekentosj&SrcApp=Papers&DestLinkType=FullRecord&DestApp=WOS&KeyUT=A1992JX28700007}

\bibitem{Corwin2011}
Corwin I 2012 {\em Random Matrices: Theory and Applications\/} {\bf 01} 1130001

\bibitem{Ferrari2005}
Ferrari P~L and Spohn H 2005 {\em Journal of Physics A: Mathematical and
  General\/} {\bf 38} L557
  \urlprefix\url{http://stacks.iop.org/0305-4470/38/i=33/a=L02}

\bibitem{Sasamoto2005}
Sasamoto T 2005 {\em Journal of Physics A: Mathematical and General\/} {\bf 38}
  L549 \urlprefix\url{http://stacks.iop.org/0305-4470/38/i=33/a=L01}

\bibitem{Calabrese2011}
Calabrese P and Le~Doussal P 2011 {\em Phys. Rev. Lett.\/} {\bf 106}(25) 250603
  \urlprefix\url{http://link.aps.org/doi/10.1103/PhysRevLett.106.250603}

\bibitem{Derrida1993}
Derrida B, Evans M~R, Hakim V and Pasquier V 1993 {\em J. Phys. A: Math.
  Gen.\/} {\bf 26} 1493
  \urlprefix\url{http://stacks.iop.org/0305-4470/26/i=7/a=011}

\bibitem{Blythe2007}
Blythe R~A and Evans M~R 2007 {\em J. Phys. A: Math. Theor.\/} {\bf 40} R333
  \urlprefix\url{http://stacks.iop.org/1751-8121/40/i=46/a=R01}

\bibitem{MHa2002}
Ha M and den Nijs M 2002 {\em Phys. Rev. E\/} {\bf 66}(3) 036118
  \urlprefix\url{http://link.aps.org/doi/10.1103/PhysRevE.66.036118}

\bibitem{Derrida1993a}
Derrida B, Janowsky S~A, Lebowitz J~L and Speer E~R 1993 {\em Journal of
  Statistical Physics\/} {\bf 73} 813--842 ISSN 1572-9613
  \urlprefix\url{http://dx.doi.org/10.1007/BF01052811}

\bibitem{Drossel2002}
Drossel B and Kardar M 2002 {\em Phys. Rev. B\/} {\bf 66}(19) 195414
  \urlprefix\url{http://link.aps.org/doi/10.1103/PhysRevB.66.195414}

\bibitem{CSChin2002}
Chin C~S 2002 {\em Phys. Rev. E\/} {\bf 66}(2) 021104
  \urlprefix\url{http://link.aps.org/doi/10.1103/PhysRevE.66.021104}

\bibitem{HKim2011}
Kim H and Huse D~A 2011 {\em Phys. Rev. B\/} {\bf 83}(5) 052405
  \urlprefix\url{http://link.aps.org/doi/10.1103/PhysRevB.83.052405}

\bibitem{Ueda2015}
Ueda M and Sasa S~i 2015 {\em Phys. Rev. Lett.\/} {\bf 115}(8) 080605
  \urlprefix\url{http://link.aps.org/doi/10.1103/PhysRevLett.115.080605}

\bibitem{MHa2011}
Ha M 2011 {\em New Physics: Sae Mulli\/} {\bf 61} 909--914 ISSN 0374-4914
  \urlprefix\url{http://www.npsm-kps.org/journal/DOIx.php?id=10.3938/NPSM.61.909}

\end{thebibliography}

\begin{appendices}
\renewcommand\thefigure{\thesection.\arabic{figure}}    
    
\section{Extra Plots for the BCSOS growth with the columnar defect}
\setcounter{figure}{0}

\begin{figure}[hb]
	\centering
	\includegraphics[width=0.9\textwidth]{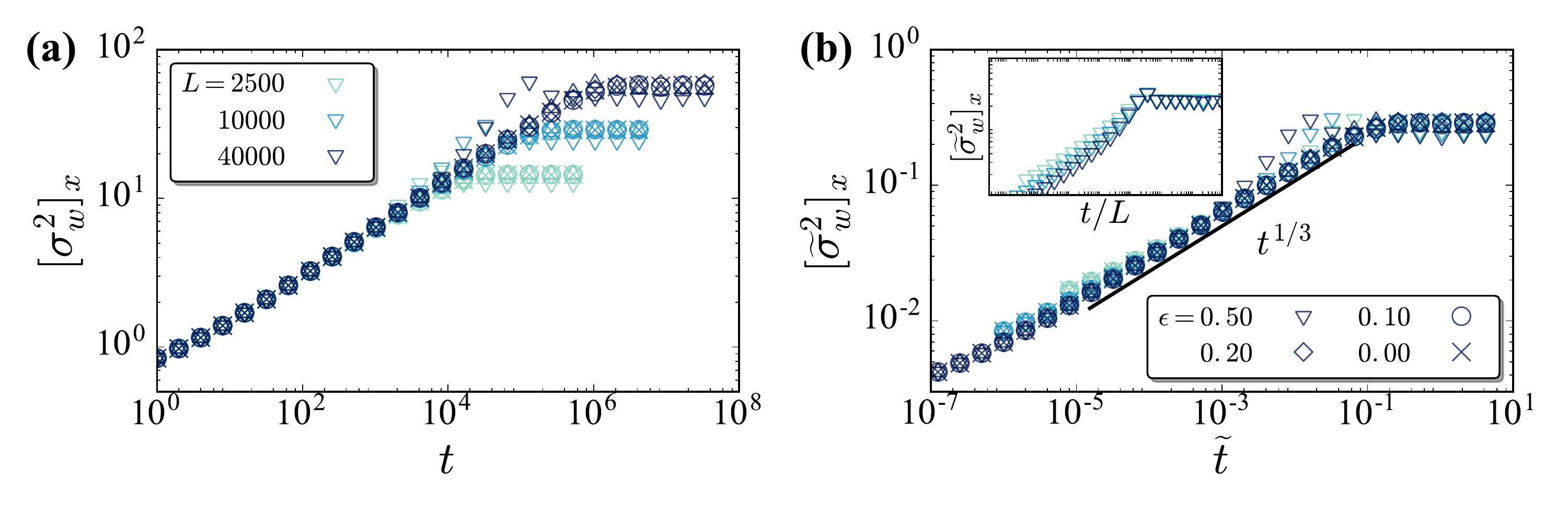}    
	\caption{Dynamic scaling of height fluctuations (related Fig.~\ref{fig:width}): $\sigma^2_w$ is defined in Eq.~(\ref{eq:sigma}). Here $\tilde{\sigma^2_w}=\sigma^2_w/L$ and $\tilde{t}=t/L^{3/2}$.} 
	\label{fig:A1}
\end{figure}

\begin{figure}
	\centering
	\includegraphics[width=0.8\textwidth]{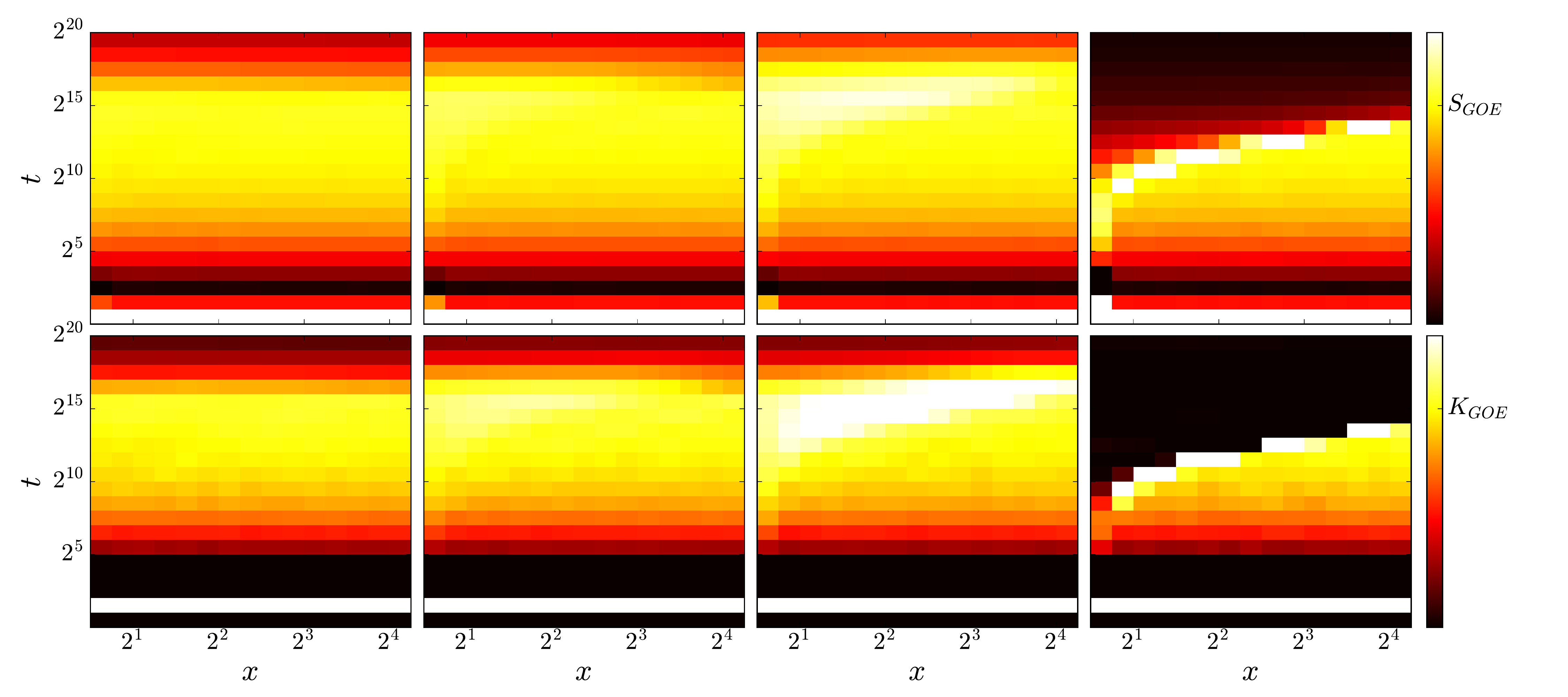}\\
        \includegraphics[width=0.8\textwidth]{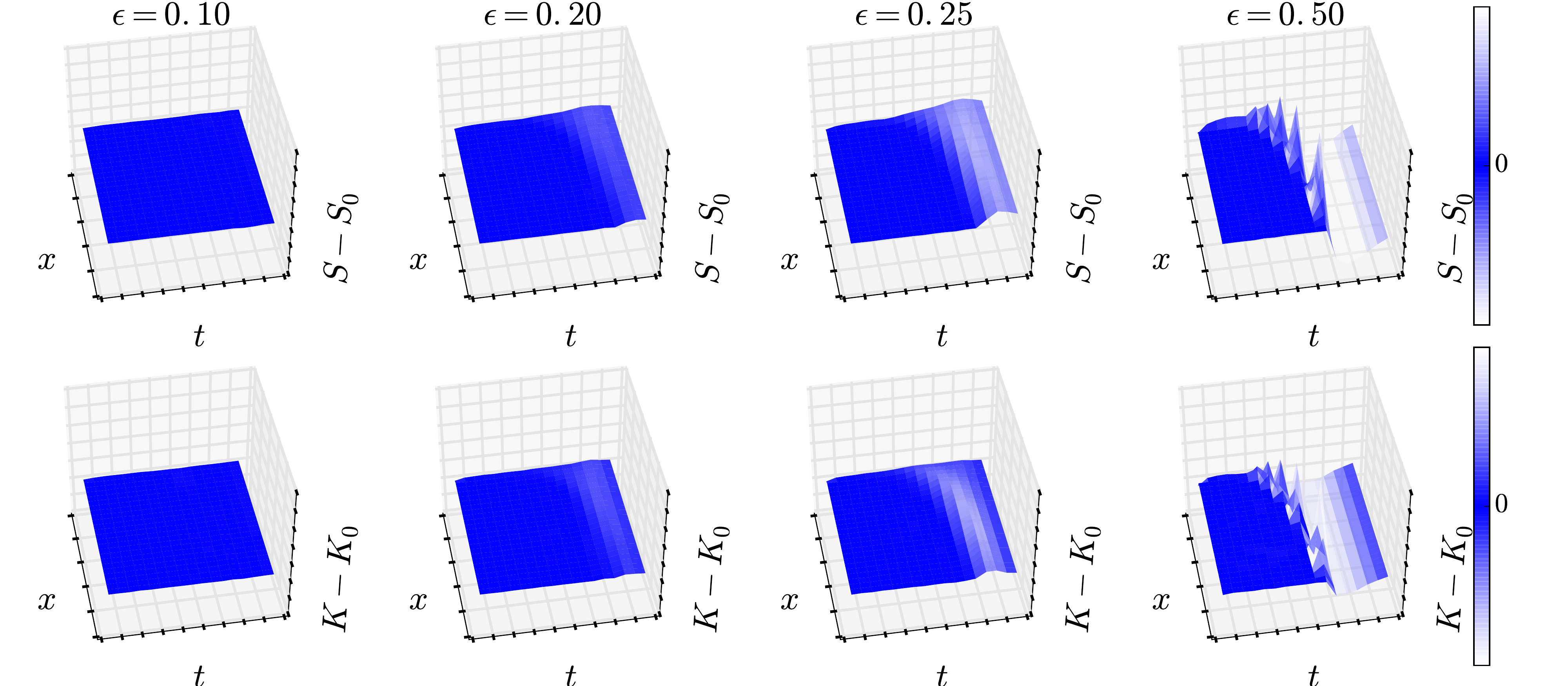}
	\caption{Skewness $S(x,t)$ and Kurtosis $K(x,t)$ (compared to Fig.~\ref{fig:3D-SnK}) are defined in Eq.~(\ref{eq:S}) and Eq.~(\ref{eq:K}), respectively. At $\epsilon=0$, we denote the skewness and the Kurtosis as $S_0$ and $K_0$, respectively.}
	\label{fig:A2}
\end{figure}

\begin{figure}
	\centering
	\includegraphics[width=0.575\textwidth]{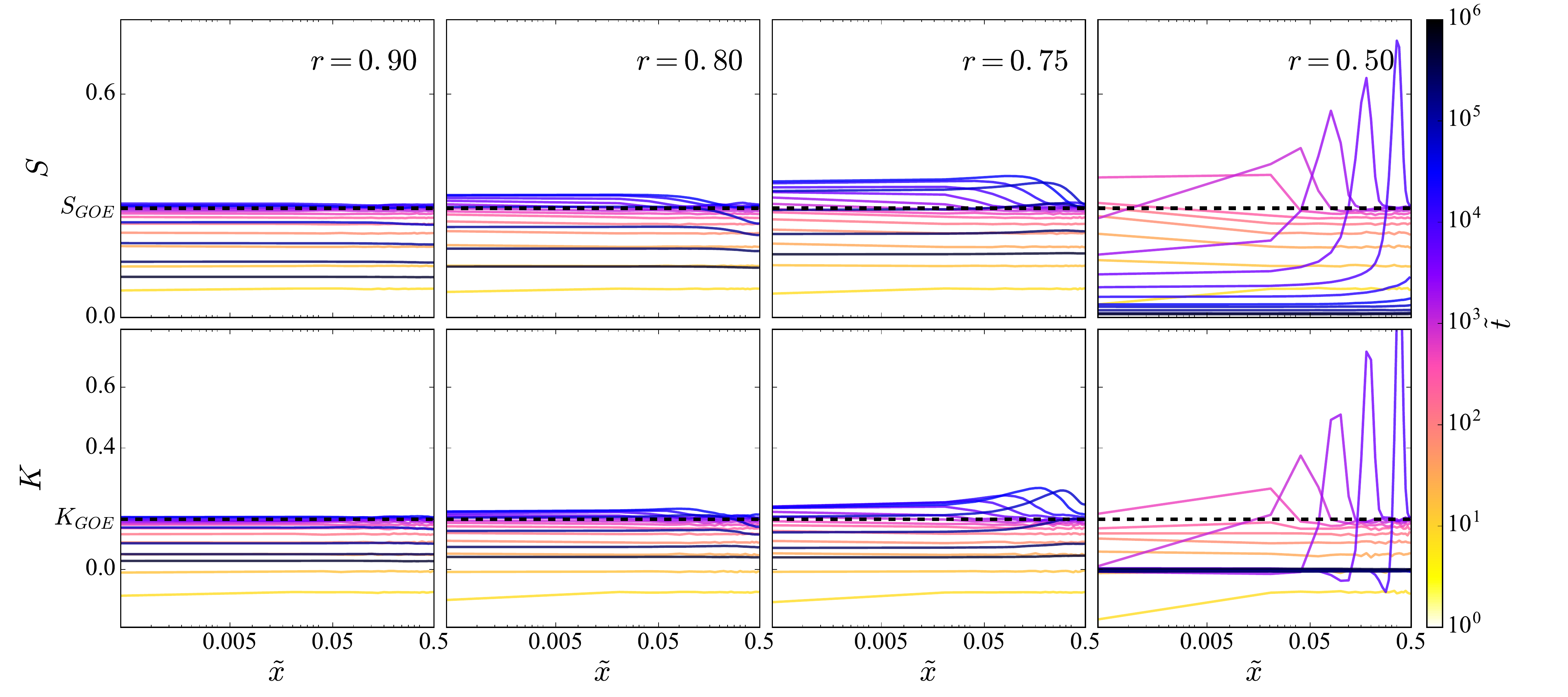} 
     \includegraphics[width=0.4\textwidth]{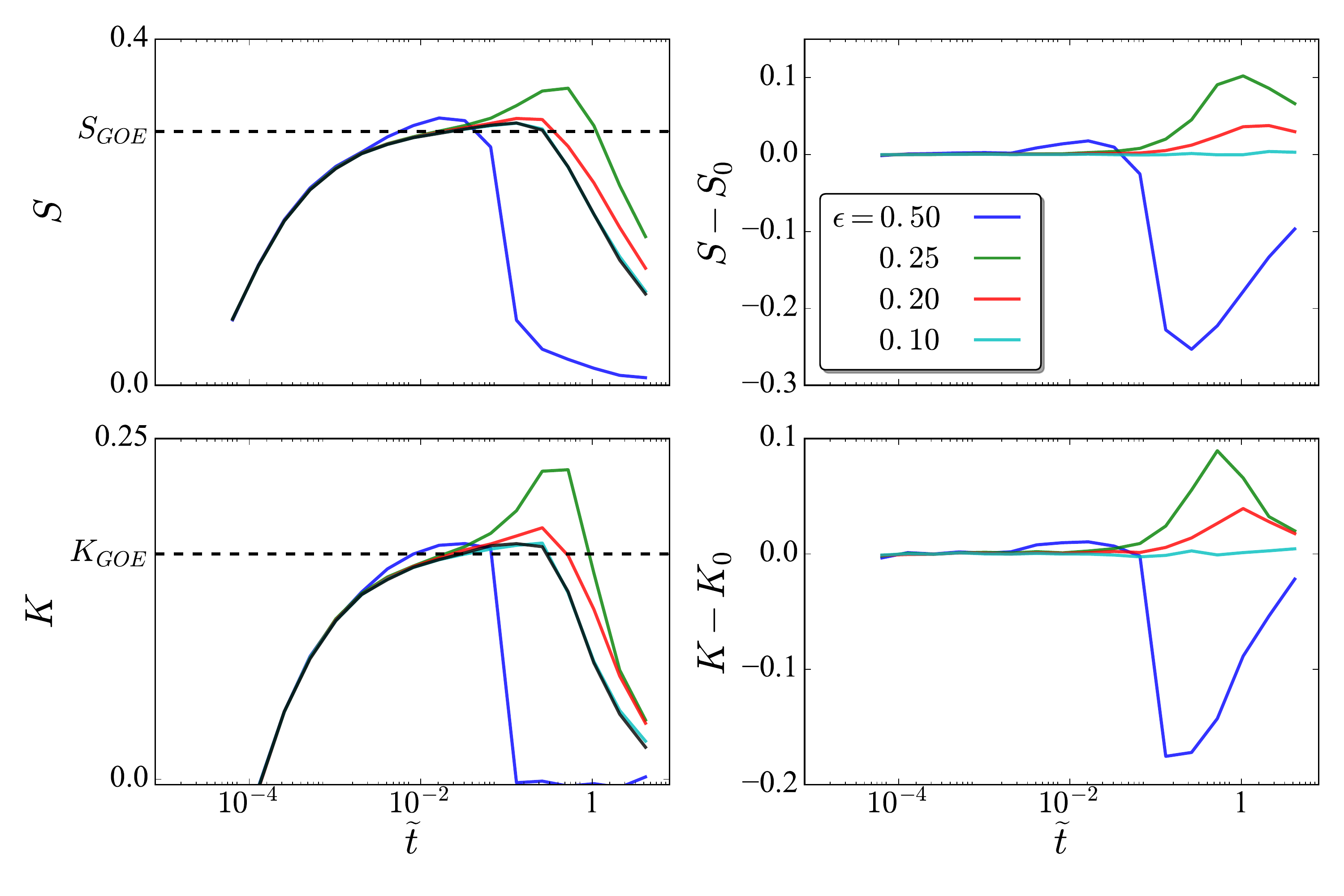}\\
        \includegraphics[width=0.575\textwidth]{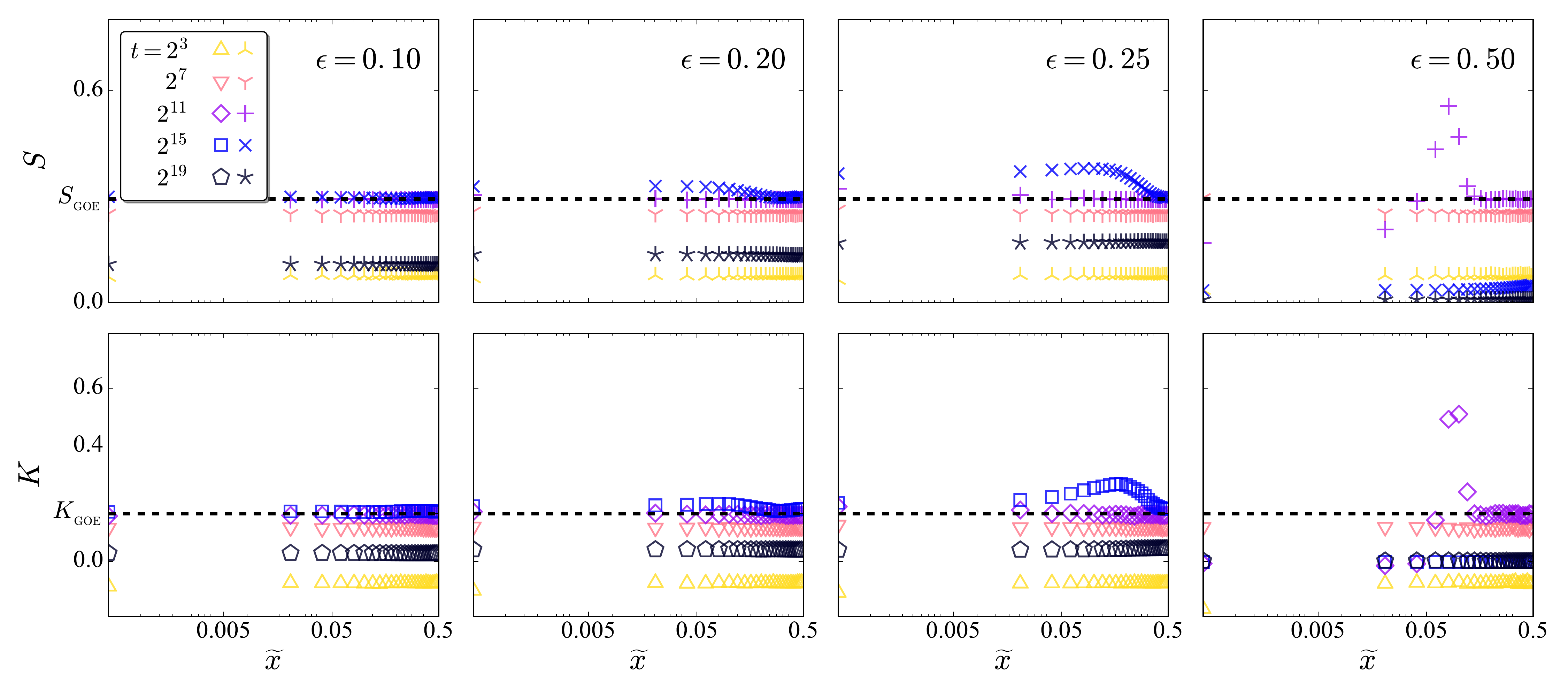} 
        \includegraphics[width=0.4\textwidth]{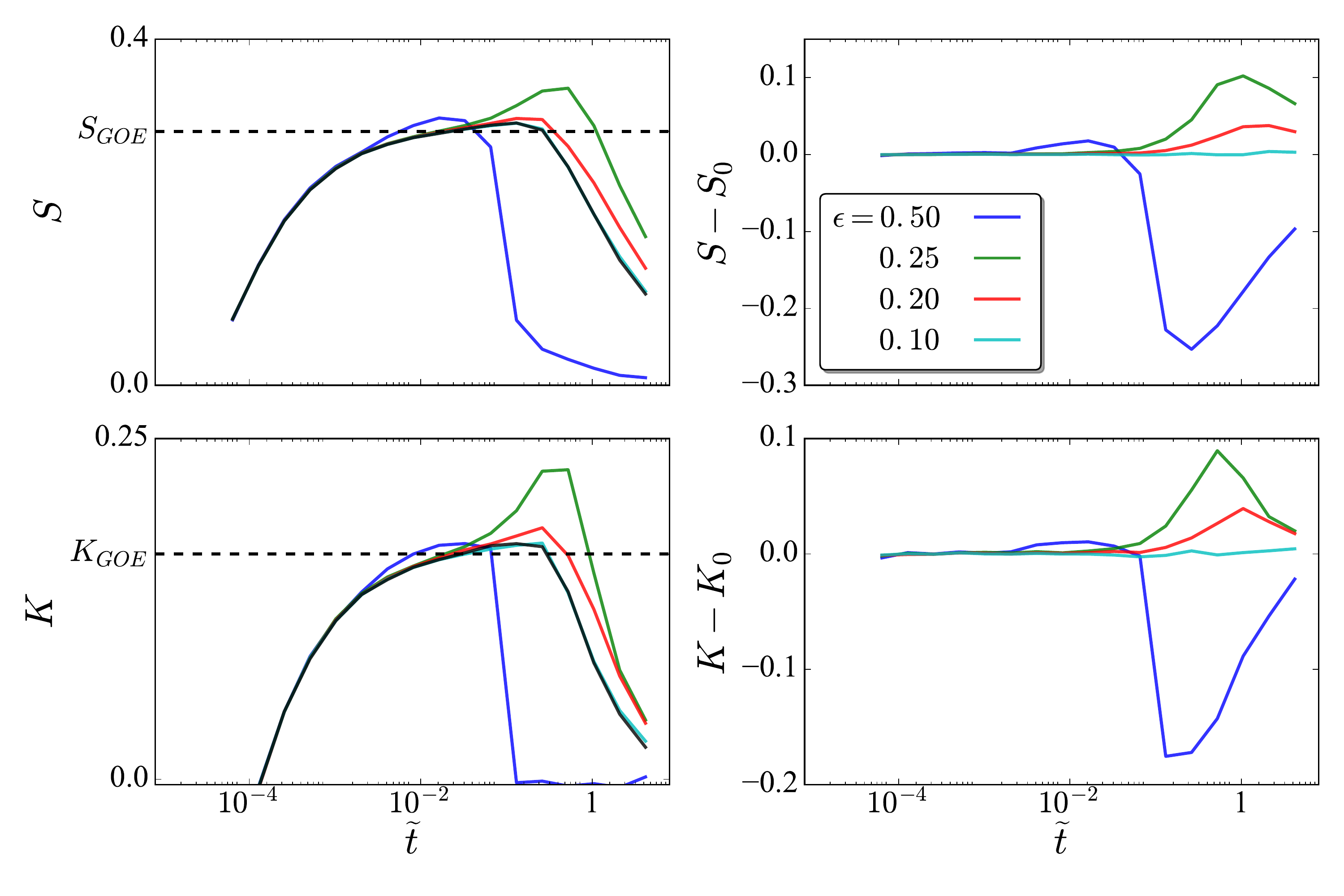}
        \includegraphics[width=0.75\textwidth]{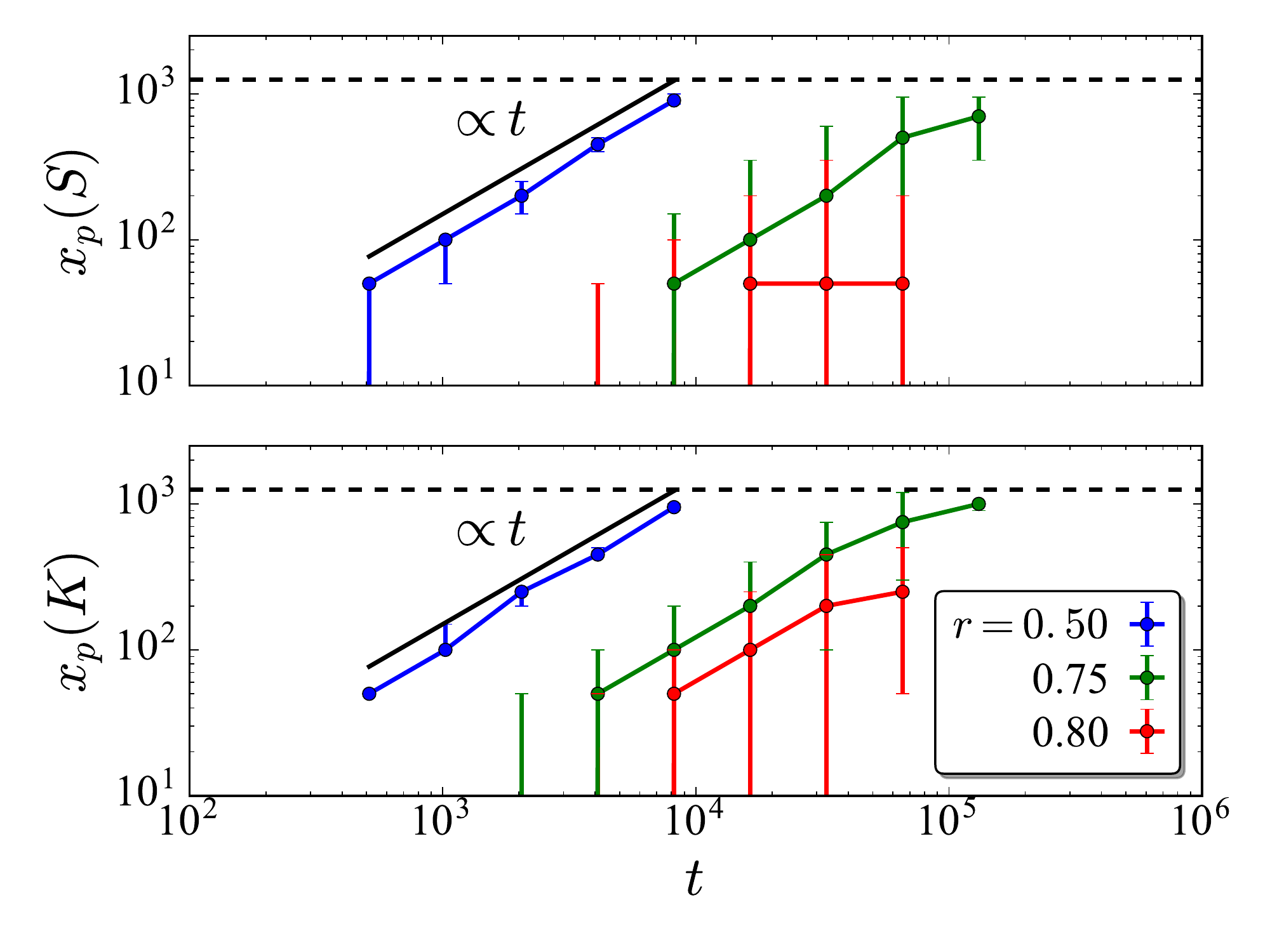}
	\caption{Skewness $S(\tilde{x},\tilde{t})$ and Kurtosis $K(\tilde{x},\tilde{t})$ (compared to Fig.~\ref{fig:2D-SnK}) are plotted against $\tilde{x}$ and $\tilde{t}$, respectively. Here $r=1-\epsilon$, $\tilde{x}=x/L$, and $\tilde{t}=t/L^{3/2}$.}
	\label{fig:A3}
\end{figure}

We here provide extra plots of asymptotic properties for the BCSOS growth with the columnar defect. The height fluctuations are plotted with the dynamic scaling in the figure of \ref{fig:A1} (a) and (b). The spatio-temporal skewness and kutorsis are plotted as heatmaps and 3-dimensional formats in figure~\ref{fig:A2}. The detailed analysis of figure~\ref{fig:A2} is provided in figure~\ref{fig:A3}.   
\end{appendices}

\end{document}